\documentclass[aip,jcp,amsmath,reprint,notitlepage,superscriptaddress,a4paper, floatfix]{revtex4-1}

\usepackage[english]{babel}
\usepackage{graphicx}
\usepackage{amssymb}
\usepackage{amsmath}
\usepackage{amsfonts}
\usepackage{color}
\usepackage[update,prepend]{epstopdf}

\newcommand{\la}{\left\langle}
\newcommand{\ra}{\right\rangle}
\def \figwidth {8.2cm}

\begin{document}
\title{Structure Formation of Amphiphilic Nanocubes at Rest and Under Shear}

\author{Takahiro Yokoyama}
\affiliation{Department of Mechanical Engineering, Keio University, 223-8522 Yokohama, Japan}

\author{Yusei Kobayashi}
\affiliation{Faculty of Mechanical Engineering, Kyoto Institute of Technology, Matsugasaki, Sakyo-ku, Kyoto 606-8585, Japan}

\author{Noriyoshi Arai}
\affiliation{Department of Mechanical Engineering, Keio University, 223-8522 Yokohama, Japan}

\author{Arash Nikoubashman}
\email{anikouba@uni-mainz.de}
\affiliation{Institute of Physics, Johannes Gutenberg University Mainz, Staudingerweg 7, 55128 Mainz, Germany}
\affiliation{Department of Mechanical Engineering, Keio University, 223-8522 Yokohama, Japan}

\date{\today}

\begin{abstract}
We investigate the self-assembly of amphiphilic nanocubes under rest and shear using molecular dynamics (MD) simulations and kinetic Monte Carlo (KMC) calculations. These particles combine both interaction and shape anisotropy, making them valuable models for studying folded proteins and DNA-functionalized nanoparticles. The nanocubes can self-assemble into various finite-sized aggregates ranging from rods to self-avoiding random walks, depending on the number and placement of the hydrophobic faces. Our study focuses on suspensions containing multi- and one-patch cubes, with their ratio systematically varied. When the binding energy is comparable to the thermal energy, the aggregates consist of only few cubes that spontaneously associate/dissociate. However, highly stable aggregates emerge when the binding energy exceeds the thermal energy. Generally, the mean aggregation number of the self-assembled clusters increases with the number of hydrophobic faces and decreases with the fraction of one-patch cubes. In sheared suspensions, the more frequent collisions between nanocube clusters lead to faster aggregation dynamics but also to smaller terminal steady-state mean cluster sizes. The MD and KMC simulations are in excellent agreement, and the analysis of the rate kernels enables the identification of the primary mechanisms responsible for the (shear-induced) cluster growth and breakup.
\end{abstract}

\maketitle

\section{Introduction}
The assembly of nanoparticles into superstructures can be directed by patterning their surface with small functional ``patches'',\cite{antonietti:ac:1997, pawar:mrc:2010, glotzer:nm:2007, li:csr:2020} thereby introducing highly localized interactions with controlled valence. In contrast to isotropic nanoparticles, the interactions between patchy particles are highly directional, which enables the formation of finite-sized aggregates such as micelles and vesicles\cite{sciortino:prl:2009, bianchi:sm:2015, kobayashi:sm:2016, kobayashi:sm:2020, li:csr:2020} as well as system-spanning gel-like networks\cite{bianchi:prl:2006, sciortino:cocis:2017} at thermodynamic equilibrium. This large spectrum of feasible superstructures and the ability to tune the (surface) chemistry of the individual particles facilitates a wide range of applications, including bio-imaging and targeted drug delivery,\cite{tran:eodd:2014, su:mtb:2019} interfacial stabilization,\cite{bradley:cocis:2017, morozova:lng:2019, morozova:acscs:2020, correia:nm:2021} and catalysis.\cite{kirillova:acsami:2015, marschelke:cps:2020}

The self-assembly behavior of nanoparticles also depends on their shape,\cite{glotzer:nm:2007} thus providing an additional handle for controlling their structure formation. Natural particles cover a diverse range of shapes, including rod-like tobacco mosaic viruses\cite{beijerinck:vka:1898, bawden:nat:1936} and gibbsite platelets,\cite{kooij:nat:2000} which have inspired the development of experimental synthesis methods that can produce a variety of non-spherical nanoparticles.\cite{glotzer:nm:2007, pawar:mrc:2010, li:csr:2020} Among these, cubic nanoparticles are particularly interesting due to their simple non-spherical geometry and space-filling properties.\cite{john:jcp:2004, agarwal:nm:2011, damasceno:sci:2012, smallenburg:pnas:2012} Nanometer-sized cubes exhibit interesting material properties, which can differ strongly from their bulk counterparts because of finite-size and surface effects. For example, hematite nanocubes demonstrate superparamagnetic performance at room temperature,\cite{wang:jpcc:2007} while Au/Ag/Au core-shell-shell nanocubes combine the strong plasmonic properties of silver with the stable and functional surface chemistry of gold.\cite{mayer:ac:2017} Perovskites based on, e.g., Cs-PB-X, can also form cubic nanocrystals, which exhibit outstanding optical properties such as color-pure photoluminescence, making them promising low-cost optoelectronics components.\cite{toso:acs:2021} 

The combination of shape and interaction anisotropy provides a powerful handle to control self-assembly and the resulting materials properties.\cite{glotzer:nm:2007, kobayashi:lng:2022, zhang:jcp:2023, argun:arxiv:2023} However, in practice it is often challenging to selectively modify the surfaces of nanocubes using bottom-up strategies, as typically all sides have the same surface chemistry. One promising pathway for creating nanocubes with patterned surfaces is DNA origami,\cite{rothemund:nat:2006, seeman:nr:2017, dey:nr:2021} as demonstrated experimentally by Scheible {\it et al.}\cite{scheible:small:2015} Alternatively, the sequence of a protein can be designed such that it folds into a cube-shaped tertiary structure with highly specific interactions on its exposed surfaces, thus resembling a patchy particle;\cite{liu:jcp:2007, mcmanus:coc:2016} for example, Du {\it et al.} recently engineered L-rhamnulose-1-phosphate proteins with histidine-mediated interactions, which fold into cube-shaped building blocks that can spontaneously self-assemble into supramolecular structures like nanoribbons or double-helical structures, depending on the pH of the buffer solution.\cite{du:nl:2021}

Evidently, nanocubes feature a rich self-assembly behavior, which depends on, {\it e.g.}, their concentration and inter-particle interactions. Systematically exploring this vast parameter space is, however, challenging for experiments alone, as it requires the time- and resource-intensive particle synthesis variation of many parameters. Computer simulations are ideally suited for this task, as they provide direct control over the particle properties and allow for efficiently exploring parameter space. In a recent article,\cite{kobayashi:lng:2022} we studied through simulations the equilibrium self-assembly of amphiphilic cubes in the limit of non-reversible aggregation, finding elongated rod-like structures and tightly packed aggregates, depending on the number and arrangement of the hydrophobic faces on the cubes. In this work, we investigate in more detail the structure formation of multi-patch cubes at finite temperatures, finding extremely slow self-assembly dynamics in some cases due to the gradual recombination of the aggregates. We further simulate suspensions under shear flow, which exhibit accelerated assembly dynamics and shear-induced breakup at sufficiently high shear rates.

\section{Model and Methods}
\label{sec:model}
\begin{figure}[htb]
    \centering
    \includegraphics[width=6cm]{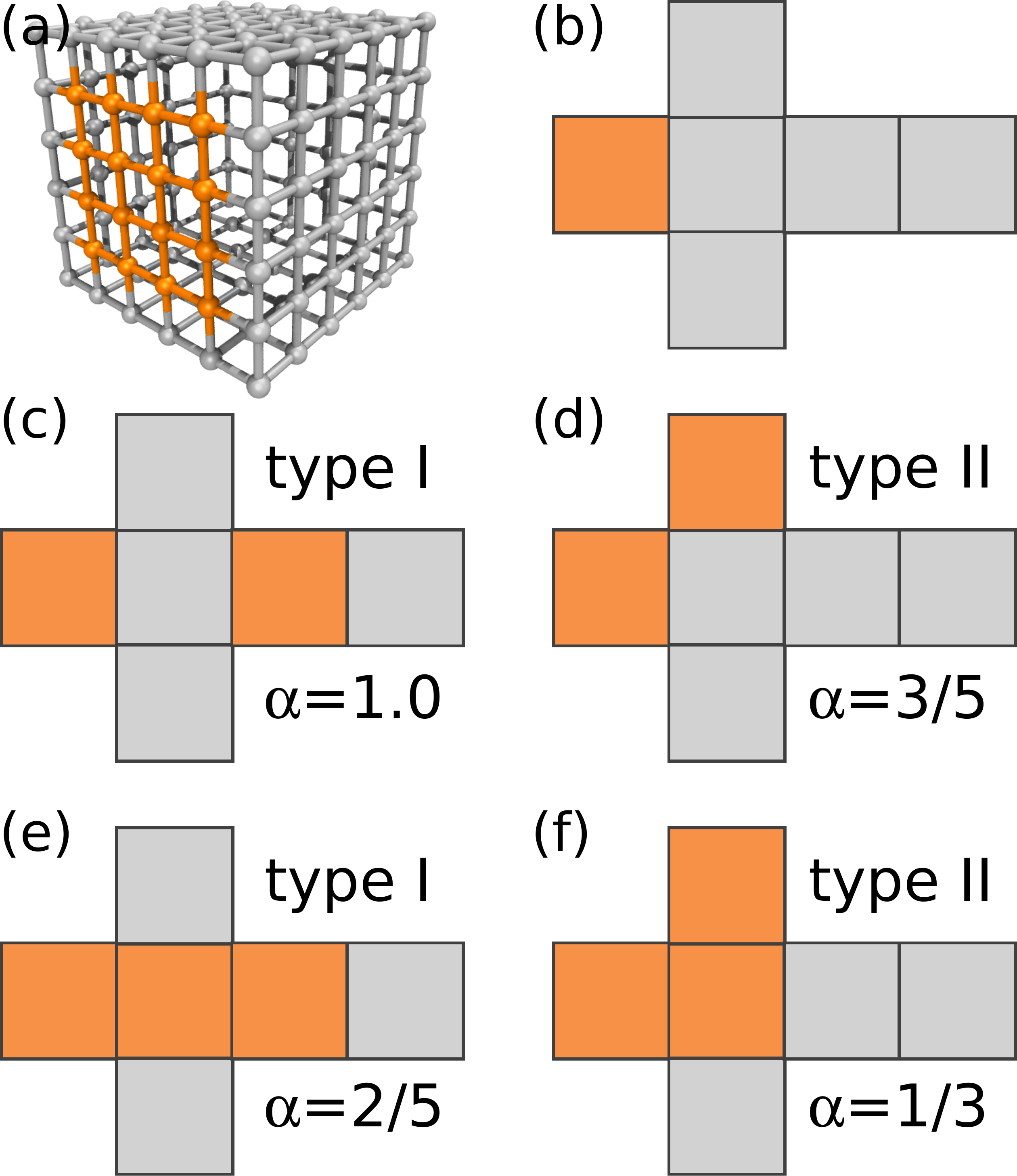}
    \caption{(a) Discrete particle model of a terminal one-patch nanocube with diameter $d=5\,a_\text{v}$. Vertex particles and bonds between nearest neighbors are shown, whereas diametric bonds have been omitted for clarity. (b-f) Unfolded representations of one-patch, two-patch, and three-patch cubes, including scaling exponent $\alpha$ used in the KMC simulations. In all panels, hydrophilic patches are colored in gray and hydrophobic ones in orange. Snapshot rendered using Visual Molecular Dynamics (version 1.9.3).\cite{vmd}}
    \label{fig:model}
\end{figure}

We specifically investigate binary mixtures of amphiphilic nanocubes, where a fraction $f$ of the cubes have only one hydrophobic face, while the remaining cubes have either two or three hydrophobic faces. Although there are 6, 15, and 20 different possibilities to place the hydrophobic patches on the one-, two-, and three-patch cubes, respectively, only a small subset of patch arrangements are geometrically distinct, as shown in Fig.~\ref{fig:model}.\cite{kobayashi:lng:2022} In the following, we refer to the two different realizations as type I and type II, respectively, as indicated in Fig.~\ref{fig:model}. We consider monodisperse suspensions, where all cubes have the same edge length $d$. This simplification is reasonable given that experimentally synthesized nanocubes\cite{ma:acsnano:2010, royer:sm:2015, sajjadi:env:2017, mayer:ac:2017} and folded proteins\cite{du:nl:2021} typically have narrow size distributions. To examine the self-assembly of the nanocubes, we use molecular dynamics (MD) simulations with a discrete mesh model and  rejection-free Kinetic Monte Carlo (KMC) calculations. 

\subsection{Molecular Simulation}
\label{sec:model.MD}
The MD simulations use a discrete particle model for the nanocubes,\cite{poblete:pre:2014, wani:jcp:2022, kobayashi:lng:2022} which consist of $N_\text{v}$ vertex particles of diameter $a_\text{v}$ and mass $m_\text{v}$, arranged on a quadratic lattice on the cube surface [see Fig.~\ref{fig:model}(a)]. The vertex particles are connected with their nearest neighbor and with their diametrically opposite counterpart through a stiff harmonic potential to maintain a (nearly) rigid shape. We chose a cube diameter of $d = 5\,a_\text{v}$ with lattice spacing $5/6\,a_\text{v}$, resulting in $N_\text{v} = 152$ vertex particles per nanocube. The mass of a nanocube is then $m = N_\text{v} m_\text{v}$. Each hydrophobic face consists of 16 vertex particles [see Fig.~\ref{fig:model}(a)], and the solvent-mediated attraction between them is modeled through a pairwise Lennard-Jones (LJ) potential acting on the hydrophobic vertex particles
\begin{equation}
    U_\text{LJ}(r) = 
    \begin{cases}
    4\varepsilon\left[\left(a_\text{v}/r\right)^{12} - \left(a_\text{v}/r\right)^6 \right] ,& r \leq r_\text{cut} \\
    0, & r > r_\text{cut}
    \end{cases},
    \label{eq:ULJ}
\end{equation}
with interaction strength $\varepsilon$ and cutoff radius $r_\text{cut} = 3\,a_\text{v}$. Excluded volume interactions between hydrophilic vertex particles and between hydrophilic and hydrophobic vertex particles are modeled using the purely repulsive Weeks-Chandler-Andersen (WCA) potential\cite{weeks:jcp:1971}
\begin{equation}
    U(r) = 
    \begin{cases}
    U_\text{LJ}(r) + \varepsilon,& r \leq 2^{1/6}\,a_\text{v} \\
    0, & r > 2^{1/6}\,a_\text{v}
    \end{cases} .
    \label{eq:UWCA}
\end{equation}

Initial configurations are generated by placing $N$ randomly oriented nanocubes at random positions without overlap in a cubic simulation box with edge length $80\,a_\text{v}=16\,d$ and volume $V$. The MD simulations are performed at volume fractions below the freezing transition of hard cubes\cite{agarwal:nm:2011} at $\phi \equiv Nd^3/V = 0.014$ (58), $0.028$ (115), $0.056$ (230), and $0.112$ (459), where the numbers in parentheses denote the total number of nanocubes in the simulation box. The equations of motion are solved using a velocity Verlet integration scheme with time step $\Delta t = 0.005\,\tau$, $\tau = \sqrt{m_\text{v}a_\text{v}^2/\varepsilon}$ being the derived unit of time. Unless stated otherwise explicitly, the temperature is fixed to $T=\varepsilon/k_\text{B}$. Equilibrium simulations were performed using an implicit solvent and a Langevin thermostat with friction coefficient $\xi = m_\text{v}/\tau$, while shear simulations were conducted with an explicit solvent using the multi-particle collision dynamics (MPCD) technique.\cite{malevanets:jcp:1999, gompper:adv:2009, howard:coce:2019} For the latter, we used the same MPCD parameters as in our previous works on sheared suspensions of Janus colloids,\cite{kobayashi:sm:2020, kobayashi:lng:2020} and generated shear flow using the reverse perturbation method,\cite{mueller-plathe:pre:1999} where an externally imposed shear stress leads to the emergence of a triangular velocity profile. To determine the measurement uncertainty of our MD data, we have conducted three independent runs per state point. All simulations are performed for at least $10^9$ time steps using the HOOMD-blue software package (v. 2.9.6).\cite{anderson:cms:2020}

\subsection{Kinetic Monte Carlo}
\label{sec:model.KMC}
\begin{figure}[htb]
    \centering
    \includegraphics[width=\figwidth]{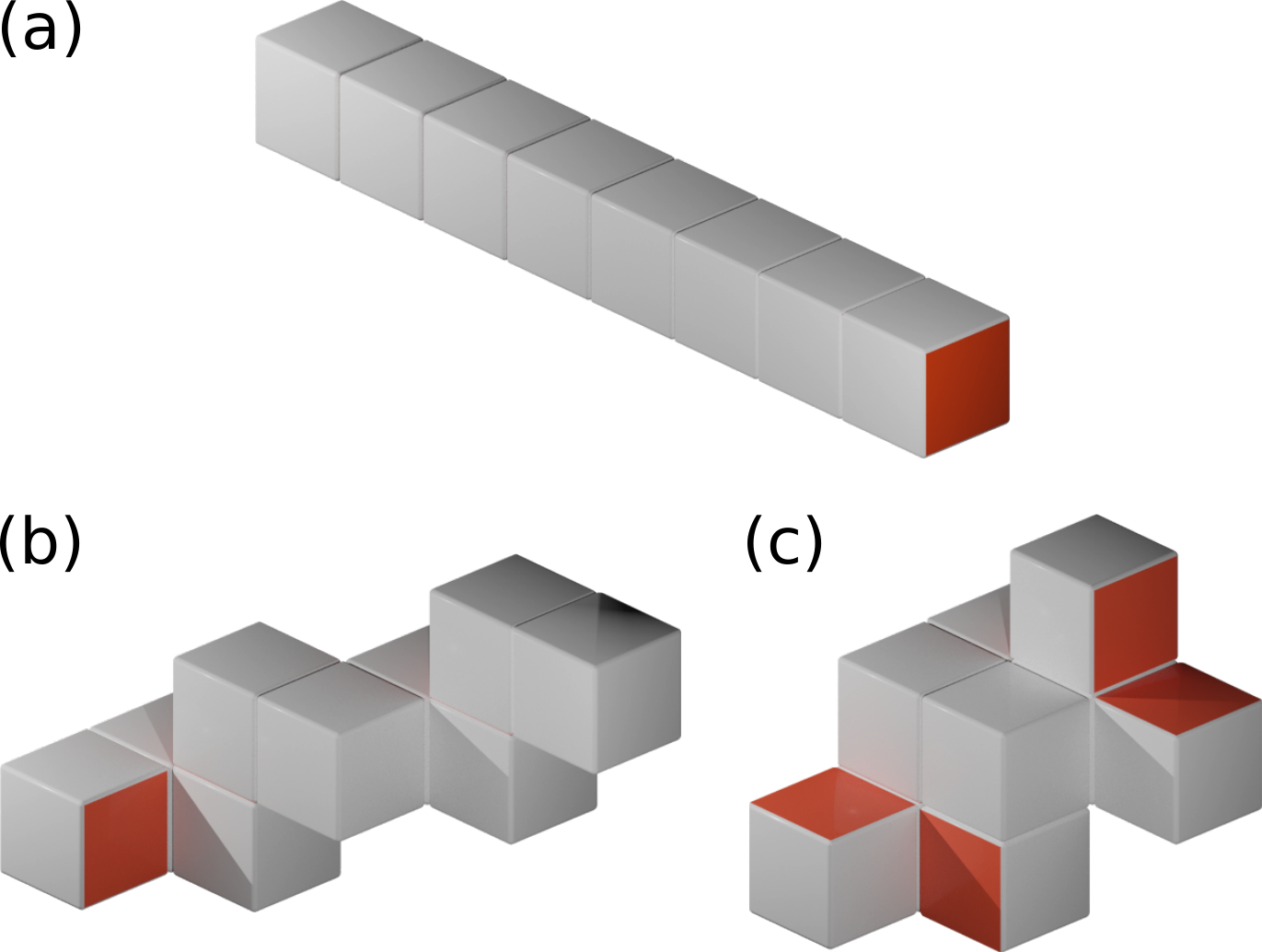}
    \caption{Schematic renderings of $M=8$ cubes self-assembled into (a) a rod ($\alpha = 1$), (b) a self-avoiding random walk ($\alpha = 3/5$), and (c) a compact configuration ($\alpha = 1/3$).}
    \label{fig:clusters}
\end{figure}
We assume that the cubes are uniformly distributed in space, and that aggregation events are pairwise. The characteristic size $L_i$ of an aggregate depends on its number of constituent cubes, $M_i$, and their arrangement (see Fig.~\ref{fig:clusters}). For example, two-patch cubes which have their hydrophobic faces on opposite sides (type I) form rod-shaped aggregates of length $L_i \simeq dM_i$, whereas the aggregated cubes follow a self-avoiding random walk with $L_i \simeq dM_i^{3/5}$ when their two hydrophobic patches lie on adjacent faces (type II).\cite{kobayashi:lng:2022} To distinguish between these different structures, we define the characteristic aggregate size as $L_i = dM_i^\alpha$ with scaling exponent $1/3 \leq \alpha \leq 1$.

\begin{figure}[htb]
    \centering
    \includegraphics[width=\figwidth]{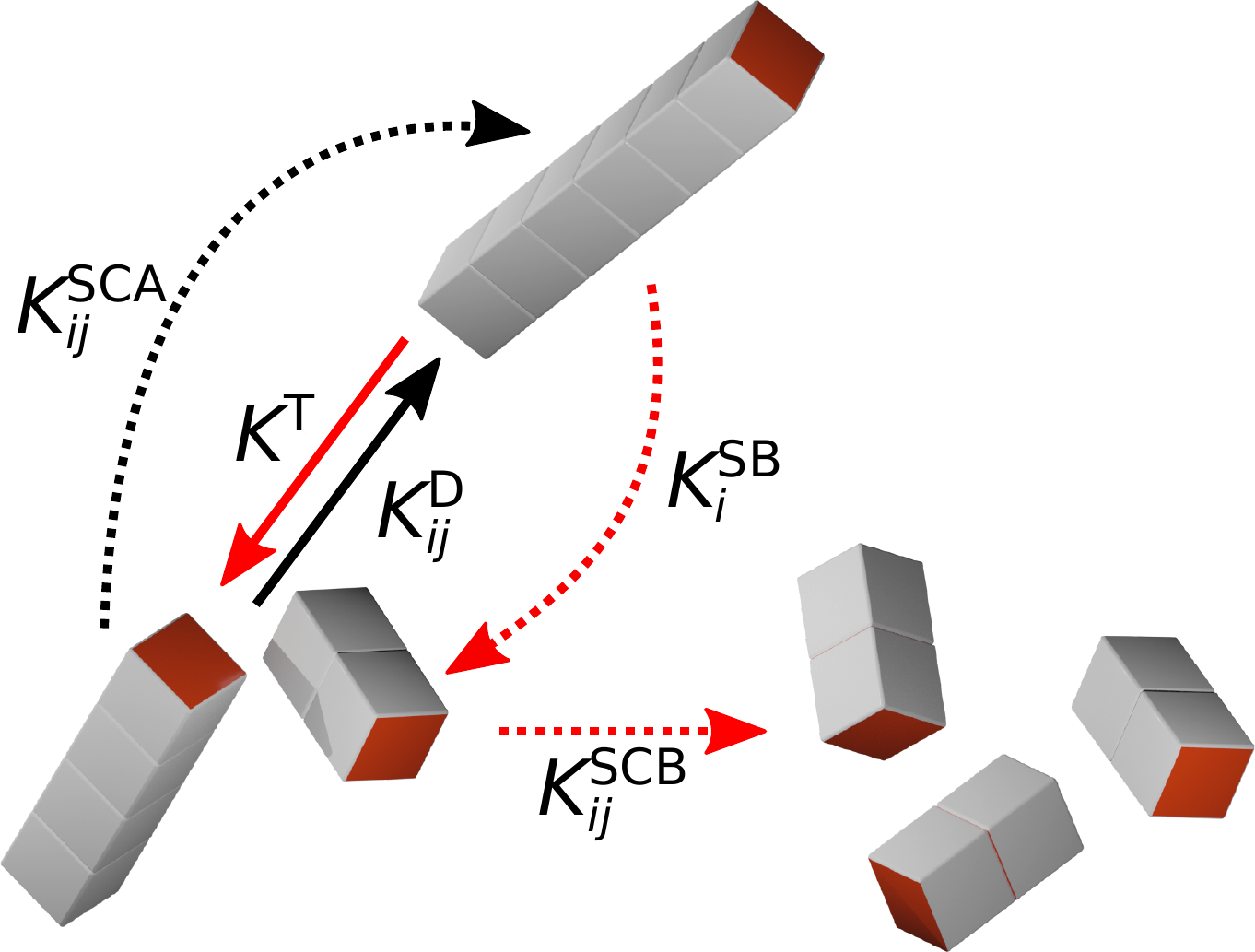}
    \caption{Schematic representation of all aggregation/breakup events. Black and red arrows indicate cluster aggregation and breakup events, respectively, while shear-induced events are indicated by dashed arrows.}
    \label{fig:kernels}
\end{figure}

\begin{figure}[htb]
    \centering
    \includegraphics[width=\figwidth]{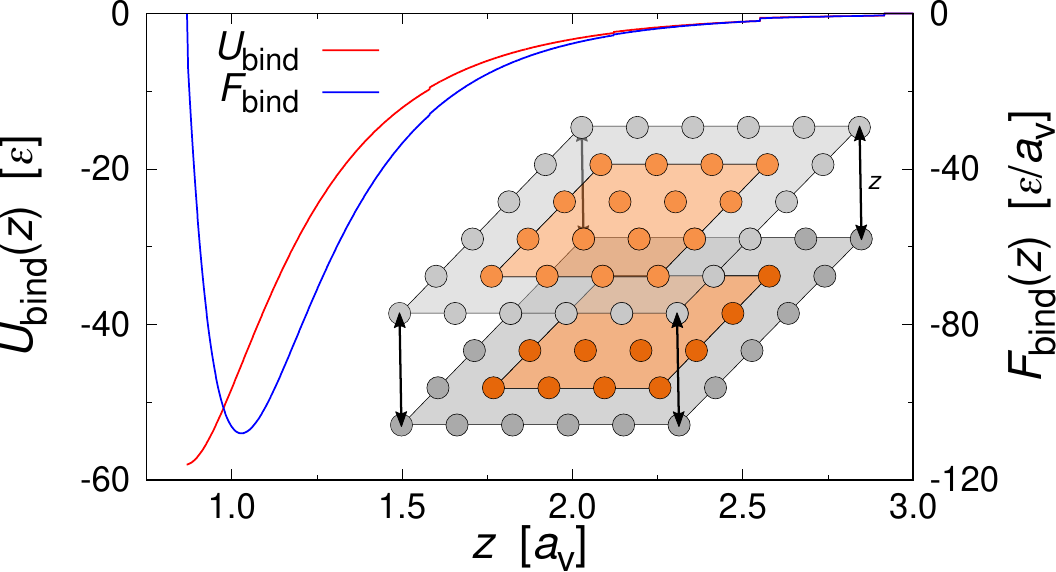}
    \caption{Binding energy $U_\text{bind}$ (left axis) and force $F_\text{bind}$ (right axis) between two discretized hydrophobic cube surfaces at distance $z$. Inset: Schematic representation of two hydrophobic cube faces in the discrete particle model at separation $z$.}
    \label{fig:patchEnergy}
\end{figure}

In KMC, the system evolves through defined processes with known transition rates. At each time step, an event is chosen based on its probability, which is calculated as the ratio of the rate of the event to the sum of the rates of all possible events. The system and transition rates are then updated, and the time is advanced using $\Delta t = -\ln(u)/K_{\rm T}$, where $u$ is a uniformly distributed random number in the interval $(0, 1]$. This procedure causes the time evolution of the system to speed up as events become less likely, in contrast to the linear time evolution of MD simulations. In our previous KMC calculations,\cite{kobayashi:lng:2022} we only considered the irreversible aggregation of one- and two-patch cubes into rod-shaped aggregates at rest. Here, we extend our KMC model to mixtures of one- and three-patch cubes and take into account five different aggregation/breakup events, which are schematically illustrated in Fig.~\ref{fig:kernels}: 
\begin{enumerate}
    \item The diffusion-driven aggregation of two clusters $i$ and $j$ is modeled using the Smoluchowski coagulation equation\cite{smoluchowski:pz:1916, smoluchowski:zpc:1918} with kernel
    \begin{equation}
        K_{ij}^\text{DA} = \frac{W_{ij}}{V} (L_i + L_j)(D_i + D_j).
        \label{eq:KDA}
    \end{equation}
    Neglecting hydrodynamic interactions and crowding effects, the translational diffusion coefficient is set to $D_i = k_{\rm B}T/(\xi M_i)$, with friction coefficient $\xi$. The parameter $W_{ij}$ controls the likelihood of two clusters $i$ and $j$ merging during a collision. In this study, we use $W_{ij} = W_iW_j$, where $W_i$ and $W_j$ are the area fraction of the exposed hydrophobic surface of clusters $i$ and $j$, respectively. For example, a single three-patch cube has $W_i = 3/6$, whereas a dimer composed of one three-patch and one one-patch cube has $W_j = 2/10$, resulting in $W_{ij} = 1/10$. Note that we erroneously wrote $K_{ij}^\text{DA} \propto 1/W_{ij}$ in our previous article,\cite{kobayashi:lng:2022} but used expression \eqref{eq:KDA} for the actual KMC simulations.
    \item Aggregates can also spontaneously break up, if the binding energy $U_\text{bind}$ is finite. Assuming dissociation into two pieces and ignoring entropic contributions to the Helmholtz free energy of the system, we model this thermally induced cluster breakup \textit{via} the kernel
    \begin{equation}
        K^\text{TB} = A\exp\left[-U_\text{bind}/(k_\text{B}T)\right] ,
        \label{eq:KTB}
    \end{equation}
    with model specific parameters $A$ and $U_\text{bind}$. To facilitate the comparison between our KMC calculations and MD simulations, we determine $U_\text{bind}$ by computing the potential energy acting between two hydrophobic faces of our discretized cube model [\textit{cf.} Fig.~\ref{fig:model}(a)]. Figure~\ref{fig:patchEnergy} shows $U_\text{bind}$ and the corresponding force $F_\text{bind}$ plotted against the separation $z$ between two hydrophobic faces. $U_\text{bind}$ and $F_\text{bind}$ are set to their minimum values of $58\,\varepsilon$ and $108\,\varepsilon/a_\text{v}$, respectively. The parameter $A=2.4 \times 10^9\,\tau^{-1}$ is chosen by matching the mean-aggregation number $\la M \ra$ in the KMC and MD simulations of two-patch particles at high temperature (see Sec.~\ref{sec:results} below).
    \item Shear can detach pieces from an isolated aggregate, if the hydrodynamic drag force exerted on the cubes exposed to the flow exceeds their characteristic binding force $F_\text{bind}$. We describe this event through the kernel\cite{icardi:tpm:2023}
    \begin{equation}
        K_i^\text{SB} = A\exp\left[-U_\text{bind}/(k_\text{B}T)\right] \frac{\xi M_i \dot{\gamma} L_i}{M_{i}^{2/3}F_\text{bind}} ,
        \label{eq:KSB}
    \end{equation}
    with shear rate $\dot{\gamma}$.
    \item In sheared systems, the clusters collide much more frequently, which can drastically affect their aggregation and breakup behavior. We incorporate shear-induced aggregation in our KMC model through the laminar shear kernel\cite{smoluchowski:zpc:1918}
    \begin{equation}
        K_{ij}^\text{SCA} = \frac{W_{ij}}{V}\dot{\gamma}(L_{i} + L_{j})^3 .
        \label{eq:KSCA}
    \end{equation}
    \item Conversely, colliding clusters can also fragment into smaller pieces if the kinetic energy of impact, $U_\text{coll}$, is comparable to the binding energy, $U_\text{bind}$. Assuming a perfectly inelastic collision, we estimate $U_\text{coll}$ as
    \begin{equation}
        U_\text{coll} = \frac{m \dot{\gamma}^2}{3}\frac{M_i M_j}{M_i + M_j}(L_i/2 + L_j/2)^2 .
    \end{equation}
    In analogy to Eq.~\eqref{eq:KSCA}, we incorporate this collision-induced breakup through the kernel
    \begin{equation}
        K_{ij}^\text{SCB} = \frac{1-W_{ij}}{V}\dot{\gamma}(L_{i} + L_{j})^3P_\text{SCB}
        \label{eq:KSCB}
    \end{equation}
    with $P_\text{SCB} = \min\left\{1, \exp\left[(U_\text{coll}-U_\text{bind})/(k_\text{B}T)\right]\right\}$.
\end{enumerate}

The total number of cubes is fixed to $N=10^4$, unless stated otherwise explicitly, and $10-100$ independent simulations were performed to gather statistics for each set of parameters. We distinguish between the different patch arrangements on the cubes through the scaling exponent $\alpha$ (see Fig.~\ref{fig:model}), which describes the shape of the self-assembled aggregates.

\section{Results}
\label{sec:results}
\subsection{Equilibrium Self-Assembly}
\begin{figure}[htb]
    \centering
    \includegraphics[width=\figwidth]{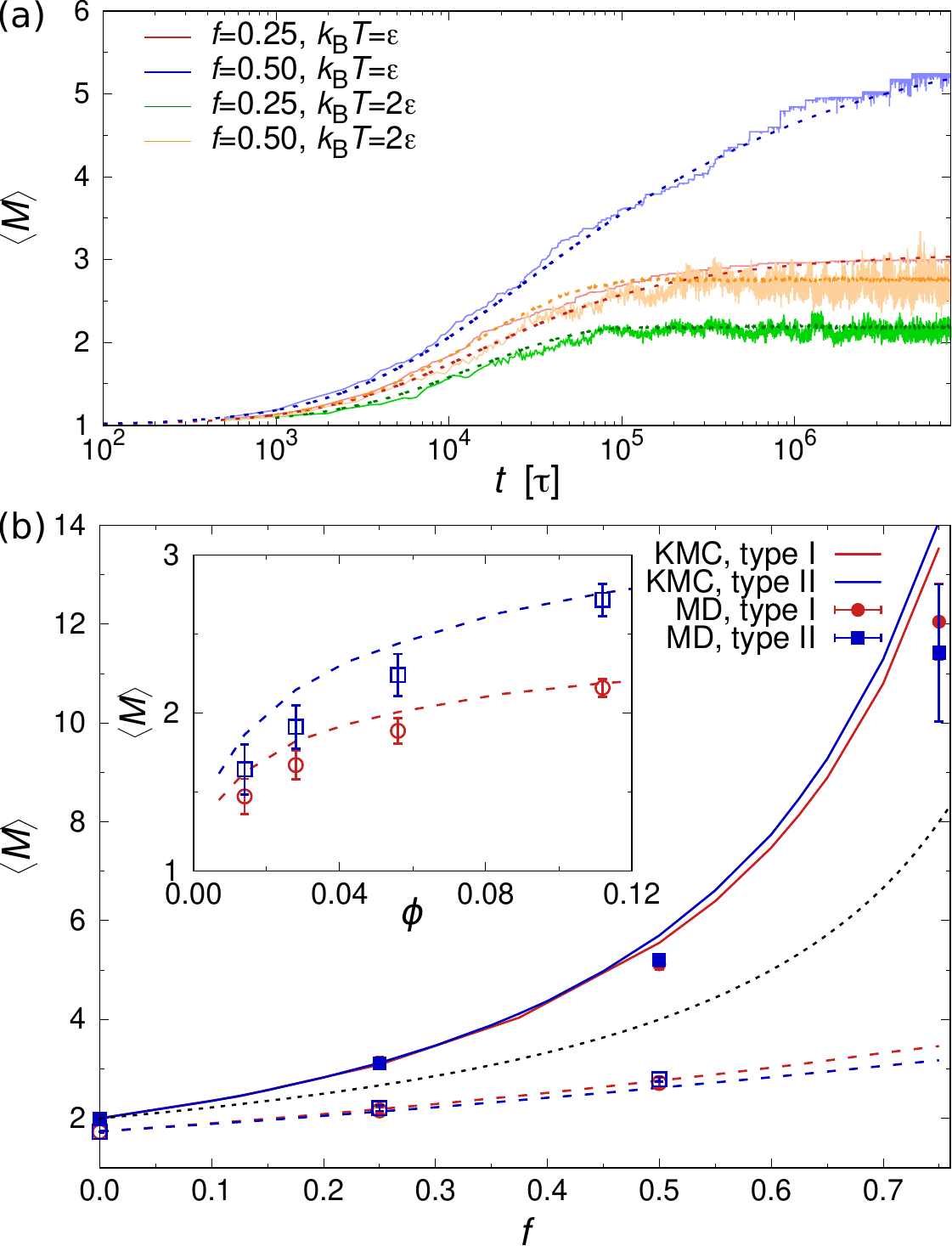}
    \caption{(a) Temporal evolution of $\la M \ra$ in mixtures of one- and two-patch cubes (type I) at $f=0.25$ and $f=0.50$ from MD (solid lines) and KMC (dashed lines) simulations for $\phi=0.112$. Data shown for temperatures $k_\text{B}T = \varepsilon$ and $k_\text{B}T = 2\,\varepsilon$. (b) Equilibrium $\la M \ra$ {\it vs} $f$ from KMC (lines) and MD (symbols) simulations at $\phi=0.112$. Solid lines and filled symbols show results for $k_\text{B}T = \varepsilon$, while dashed lines and open symbols show results for $k_\text{B}T = 2\,\varepsilon$. The black dotted line shows the expectation value $\la M \ra_\text{urn}$ of the corresponding negative hypergeometric probability distribution. Inset: $\la M \ra$ {\it vs} $\phi$ for type I case at $k_\text{B}T = 2\,\varepsilon$.}
    \label{fig:MAN2patch}
\end{figure}

Figure~\ref{fig:MAN2patch}(a) shows the mean aggregation number $\la M \ra$ as a function of time for mixtures of two- and one-patch cubes at rest. Here, $\la M \ra$ is defined as $\la M \ra \equiv \sum_i^N MP(M)$, where $P(M)$ is the probability to find a cube in an aggregate consisting of $M$ nanoparticles (see Ref.~\citenum{kobayashi:lng:2022} for technical details of the cluster analysis). The KMC and MD simulation results match closely when the time in the KMC simulations is rescaled by a constant factor, validating the KMC model. Starting from a completely dissociated state, $\la M \ra = 1$ initially, the mean aggregation number increased over time and eventually reaches a plateau equilibrium value that is higher for larger $f$. At higher temperatures, $\la M \ra$ decreased and shows stronger fluctuations due to continuous aggregate breakup and reformation.

The equilibrium value of $\la M \ra$ increased monotonically with increasing $f$ [Fig.~\ref{fig:MAN2patch}(b)], since there were fewer one-patch cubes that can terminate cluster growth ($\la M \ra = 2$ at most for a pure suspension of one-patch cubes). Interestingly, the type I and type II cases exhibit almost identical $\la M \ra$ values, suggesting that the placement of hydrophobic patches on nanocubes has little effect on the mean aggregation number. The self-assembly behavior of these one- and two-patch cube mixtures shares similarities with the statistics problem of drawing without replacement from an urn containing two particle types, with $N$ particles in total. In this context, the probability of $k$ successful draws (two-patch particles) until $r$ failed draws (one-patch particles) is given by a negative hypergeometric distribution, with expectation value
\begin{equation}
    \la M \ra_\text{urn} = \frac{rfN}{N(1-f)+1} + r
    \label{eq:Murn}
\end{equation}
including the $r$ failed attempts (one-patch particles). For clusters consisting of two-patch cubes, two one-patch cubes are needed to terminate growth, so that $r=2$. Further, for sufficiently large $N \gg 1$, Eq.~\eqref{eq:Murn} simplifies to $\la M \ra_\text{urn} \approx 2x/(1-x) + 2$. We have plotted $\la M \ra_\text{urn}$ as a dotted line in Fig.~\ref{fig:MAN2patch}, which increases similarly with $f$ like the KMC and MD simulations, but systematically lies below the simulation results. This difference likely stems from the fact that only one particle is drawn at a time in the urn model, whereas the KMC and MD simulations allow merging events between clusters with $M_i, M_j > 1$.

At fixed $f$, $\la M \ra$ increased monotonically with increasing nanoparticle volume fraction $\phi$ in both KMC and MD simulations [inset of Fig.~\ref{fig:MAN2patch}(b)]. This behavior can be explained by recognizing that the rate of diffusion-driven aggregation between two clusters depends on the characteristic distance between them [given by $K_{ij}^\text{DA}$ in Eq.\eqref{eq:KDA}], while the thermal breakup rate of a cluster does not involve any other interaction partner and is thus independent of $\phi$ [given by $K^\text{TB}$ in Eq.\eqref{eq:KTB}]. Notably, our previous KMC model\cite{kobayashi:lng:2022} did not capture this $\phi$-dependence of $\la M \ra$, as it did not include any cluster breakup.

\begin{figure}[htb]
    \centering
    \includegraphics[width=\figwidth]{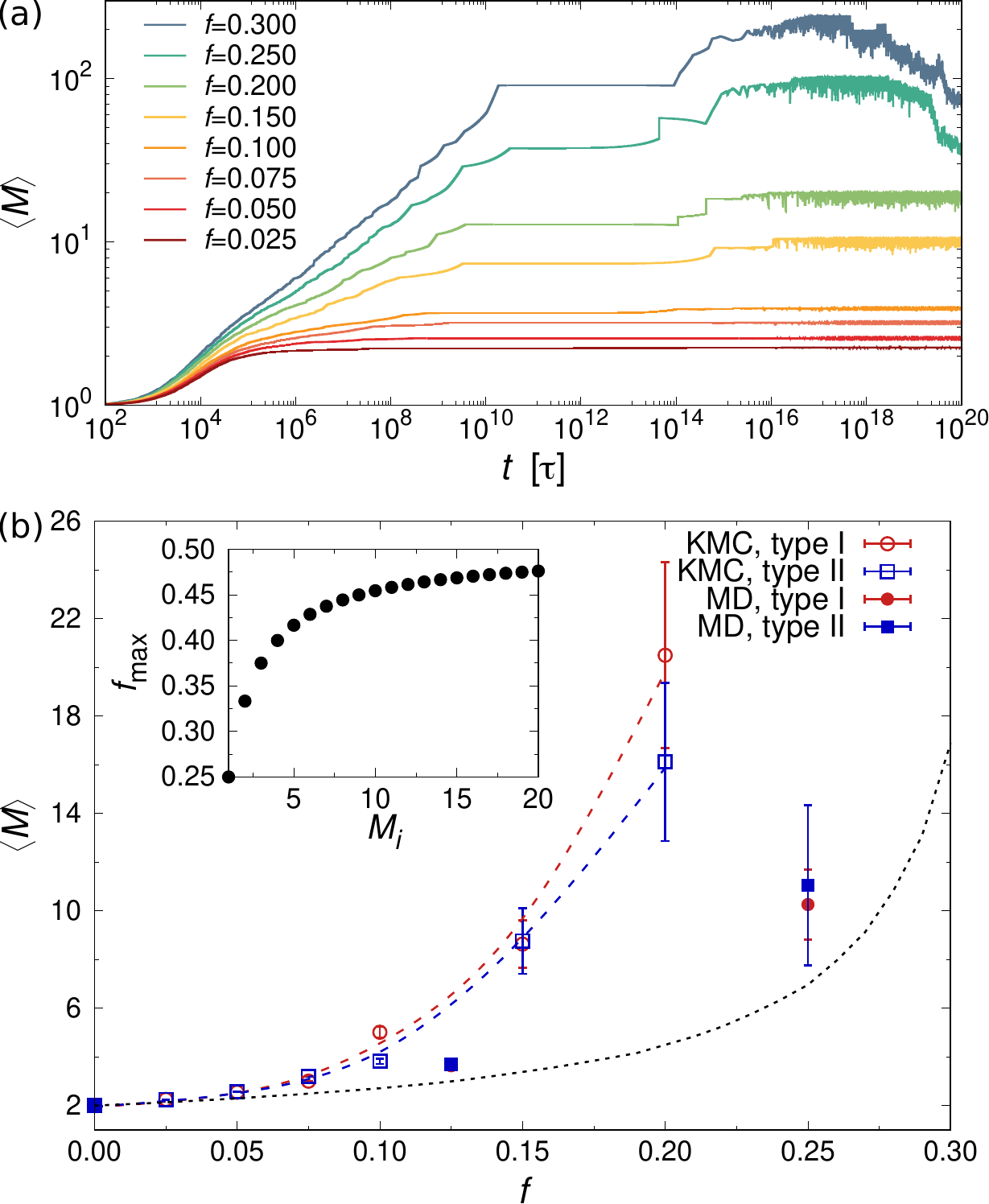}
    \caption{(a) Temporal evolution of $\la M \ra$ in mixtures of one- and three-patch cubes (type II) at $\phi=0.112$. (b) Equilibrium $\la M \ra$ {\it vs} $f$ from KMC (lines) and MD (symbols) simulations for mixtures of one- and three-patch cubes (type I) at $\phi=0.112$ and $k_\text{B}T = \varepsilon$. KMC results shown for various total number of cubes $N$, as indicated. Inset: Theoretical maximum value of $f$ needed for terminating growth of a cluster consisting of $M_i$ three-patch cubes.}
    \label{fig:MAN3patch}
\end{figure}

Next, we examined the self-assembly behavior of mixtures containing one- and three-patch cubes at rest. The temporal evolution of the mean aggregation number $\la M \ra$ is plotted in Fig.~\ref{fig:MAN3patch}(a) for mixtures containing three-patch cubes at fixed volume fraction $\phi = 0.112$. For small $f \leq 0.2$, $\la M \ra$ increased monotonically until it reached its equilibrium value. However, for larger values of $f$, $\la M \ra$ first reached an intermediate plateau before experiencing step-wise growth caused by recombination and merging of large clusters. Over longer times, the rate of thermal breakup $K^\text{TB}$ became significant, resulting in the breakup of clusters into smaller fragments, leading to a slow but steady decline in $\la M \ra$. Comparing the self-assembly dynamics of the two- and three-patch cases [{\it cf.} Figs.~\ref{fig:MAN2patch}(a) and \ref{fig:MAN3patch}(a)], we observed that the latter exhibited much slower kinetics, occurring on time scales that were more than ten orders of magnitude larger. These slow assembly dynamics highlight the challenges in equilibrating these mixtures using conventional MD simulations. Figure~\ref{fig:MAN3patch}(b) shows $\la M \ra$ as a function of the fraction $f$ of three-patch cubes obtained from the MD and KMC simulations. Like the two-patch cubes, $\la M \ra$ increased monotonically with increasing $f$, albeit the self-assembled clusters were much larger for the three-patch case at a given $f$, {\it e.g.}, $\la M \ra \approx 20$ (three-patch) {\it vs} $\la M \ra \approx 2.8$ (two-patch) in KMC simulations at $f=0.2$.

To understand the larger cluster size and slower self-assembly dynamics of three-patch cubes, it is helpful to consider the conditions for terminating cluster growth: While aggregates composed of two-patch cubes need only two one-patch cubes to stop cluster growth, irrespective of their aggregation number $M_i$, a cluster consisting of $M_i$ three-patch cubes requires $M_i + 2$ one-patch cubes to prevent further aggregation. Hence, to achieve a stable suspension of finite-sized aggregates, the fraction of multi-patch nanocubes cannot exceed $f_\text{max} \equiv M_i/(2M_i+2)$, which approaches $f_\text{max} \approx 1/2$ for large $M_i$ (see inset of Fig.~\ref{fig:MAN3patch}). In addition, the one-patch cubes must adsorb successively to the exposed hydrophobic surfaces of a three-patch cube, which has a probability of roughly $(1-f)^2$ for $M_i \geq 2$ to first approximation. Since the criteria for terminating the growth of a cluster depend on its size and composition, it is difficult to approximate the aggregation behavior using an analytically tractable urn model [{\it cf.} Eq.~\ref{eq:Murn} for the two-patch case]. We can, however, numerically compute $\la M \ra_\text{urn}$, which is shown as a dotted line in Fig.~\ref{fig:MAN3patch}.

\subsection{Shear-Induced Aggregation and Breakup}
\label{sec:shear}
\begin{figure}[htb]
    \centering
    \includegraphics[width=\figwidth]{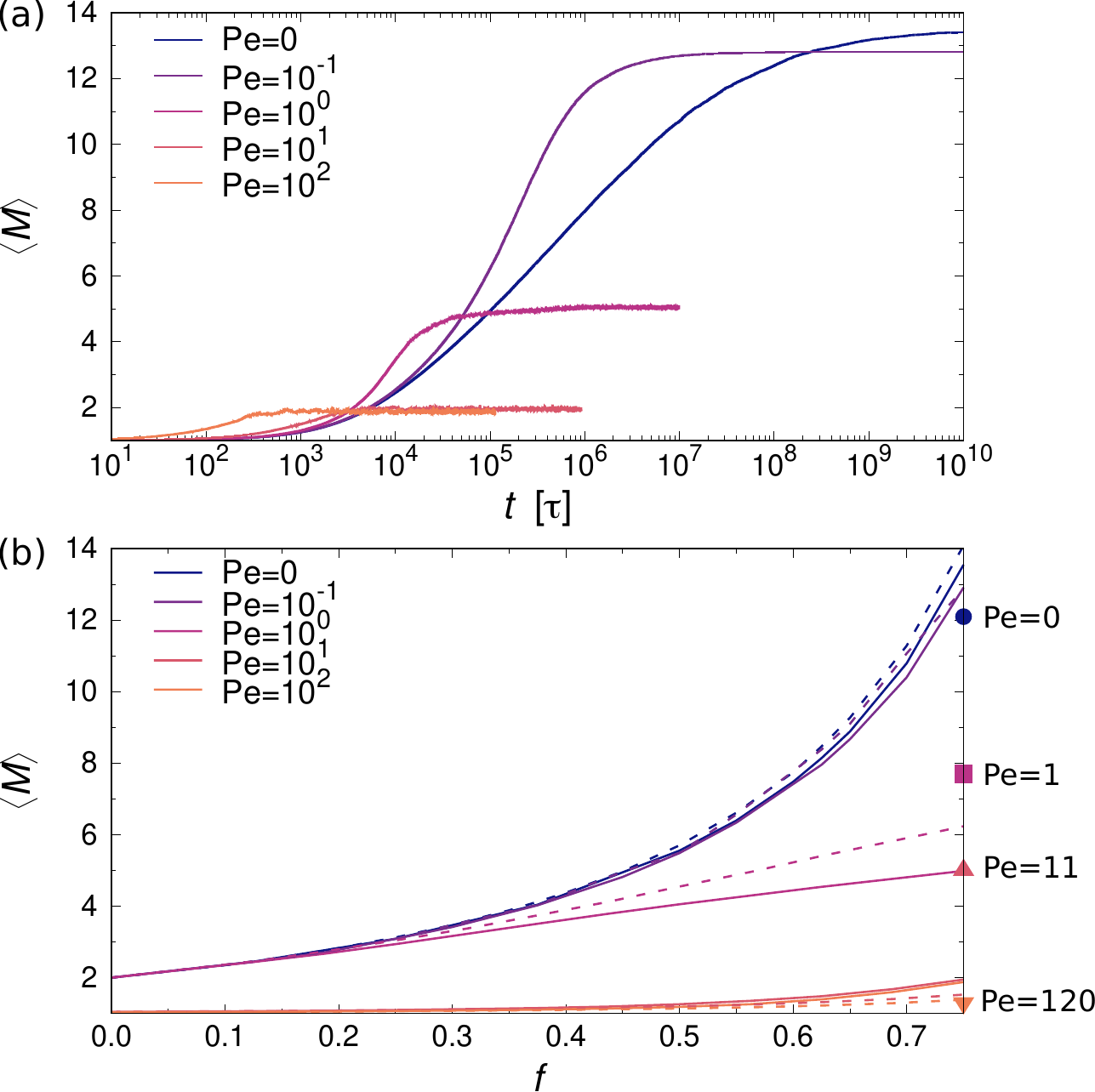}
    \caption{(a) Temporal evolution of $\la M \ra$ in mixtures of one- and two-patch cubes (type I) from KMC simulations at $f=0.75$ and $\phi=0.112$. (b) Steady-state $\la M \ra$ {\it vs} $f$ for type I (solid lines) and type II (dashed lines) two-patch cubes. Results from non-equilibrium MD simulations at $f=0.75$ and $\phi=0.112$ are shown as symbols on the right $y$-axis.}
    \label{fig:MAN2shear}
\end{figure}

Nanoparticles in sheared suspensions experience many more collisions among them compared to quiescent suspensions, which can lead to the enhanced growth or breakup of the aggregates depending on the flow conditions and particle interactions.\cite{smoluchowski:zpc:1918, oles:jcis:1992, zaccone:prl:2011, bianchi:sm:2015, kobayashi:sm:2020} In addition, shear flow can cause deformation of the aggregates, leading to changes in their shape and orientation, which can facilitate the formation of additional interparticle bonds. To characterize the strength of the shear flow, we introduce the P{\'e}clet number
\begin{equation}
    \text{Pe} \equiv \frac{\dot{\gamma}d^2}{D_0} ,
    \label{eq:Pe}
\end{equation}
which is the ratio between the rate of advection and the rate of diffusion for a single cubic particle. Figure~\ref{fig:MAN2shear}(a) shows the time evolution of $\la M \ra$ for a mixture of one- and two-patch cubes at different values of $\text{Pe}$. The data confirm that the cubes aggregate much earlier in the sheared systems, as expected from the advective transport. However, the steady-state $\la M \ra$ is significantly smaller in the sheared suspensions compared to the systems at rest. For example, at $\text{Pe} = 10^2$, the majority of aggregates are dimers, whereas at rest there are many rod-shaped aggregates consisting of more than ten cubes. These trends apply to all investigated $\phi$ and $f$, with only minor quantitative differences between the type I and type II patch arrangements, as shown in Fig.~\ref{fig:MAN2shear}(b).

For validating the KMC results for sheared suspensions, we performed non-equilibrium MD simulations at $\phi=0.112$ and $f=0.75$ (see Sec.~\ref{sec:model.MD} and Refs.~\citenum{kobayashi:sm:2020, kobayashi:lng:2020} for technical details). The resulting steady-state values of $\la M \ra$ are shown at the right $y$-axis of Fig.~\ref{fig:MAN2shear}(b), which qualitatively exhibit the same monotonic decrease of $\la M \ra$ with increasing $\text{Pe}$ as the KMC simulations, indicative of shear-induced cluster breakup. There are, however, some quantitative differences at intermediate P{\'e}clet numbers $1 \lesssim \text{Pe} \lesssim 10$, where we find systematically larger clusters in the MD simulations compared to the KMC calculations. These deviations are likely rooted in differences between our molecular and KMC models (see Sec.~\ref{sec:model}), which should become especially pronounced at intermediate $\text{Pe}$ where the aggregation and breakup kernels become comparable in magnitude, as discussed further below. 

\begin{figure}[htb]
    \centering
    \includegraphics[width=\figwidth]{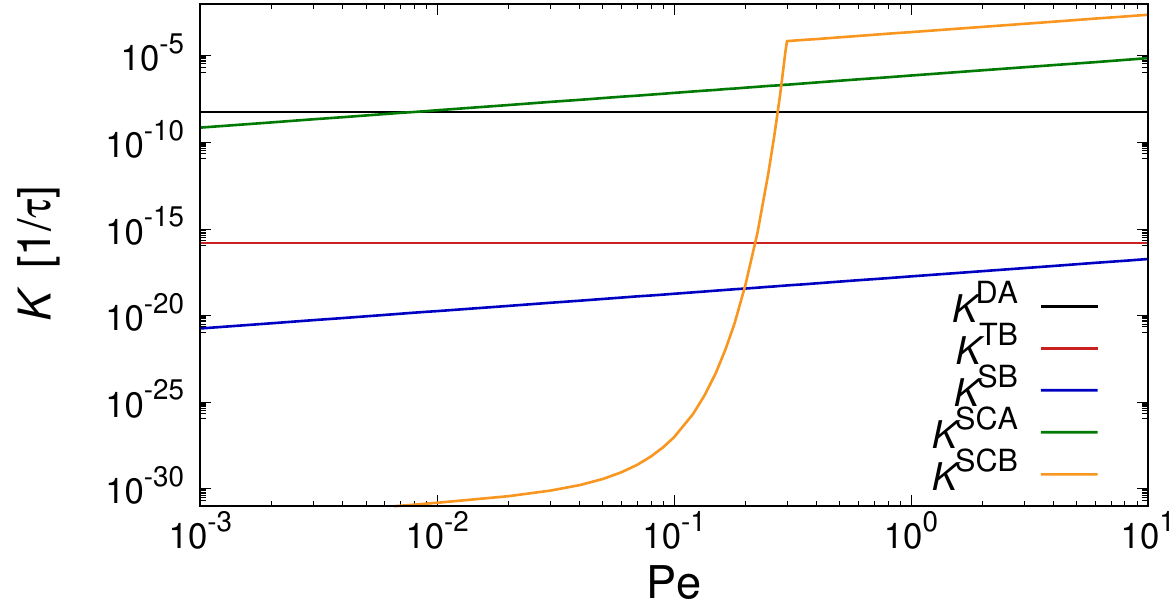}
    \caption{Rate kernels for the interactions between two aggregates with $M_i = M_j = 4$, consisting of one one-patch and three two-patch cubes (type I) at $\phi=0.112$ and $k_\text{B}T = \varepsilon$.}
    \label{fig:twoPatchKernels}
\end{figure}

In order to gain insight into the physical mechanisms driving the assembly under shear, we examined how the rate kernels $K$ change as the P{\'e}clet number increases. As an example, we have plotted in Fig.~\ref{fig:twoPatchKernels} these dependencies for the kernels describing the interactions between two aggregates with $M_i = M_j = 4$ at $k_\text{B}T = \varepsilon$. The kernels describing diffusion-driven aggregation, $K^\text{DA}$, and thermal breakup, $K^\text{TB}$, are independent of the applied shear rate $\dot{\gamma}$ [{\it cf.} Eqs.~\eqref{eq:KDA} and \eqref{eq:KTB}], with $K^\text{DA}/K^\text{TB} \sim 10^7$ at $k_\text{B}T = \varepsilon$ so that aggregation is virtually irreversible. At elevated temperature $k_\text{B}T = 2\,\varepsilon$, this ratio decreases drastically to $K^\text{DA}/K^\text{TB} \sim 10^{-5}$, thus greatly favoring the dissociation of the aggregates ({\it cf.} Fig.~\ref{fig:MAN2patch}).

To gauge the relevance of shear for the growth of the aggregates, consider the ratio $K_{ij}^\text{SCA}/K_{ij}^\text{DA}$, which becomes larger than unity when
\begin{equation}
    \text{Pe} \geq \frac{M_i^{-1} + M_j^{-1}}{(M_i+M_j)^{2\alpha}} .
    \label{eq:KSCAoverKDA}
\end{equation}
Thus, the shear-induced collision of clusters becomes the dominant aggregation mechanism already at rather small P{\'e}clet numbers, {\it i.e.}, $\text{Pe} \geq 2^{1/3}$ for the ``worst case'' of two three-patch cubes with type II patch arrangement ($M_i = M_j =1$, $\alpha = 1/3$). This threshold quickly drops with increasing aggregate size, which explains the rapid increase in $\la M \ra$ in the sheared systems.

To understand why the steady-state average cluster size decreases with increasing $\text{Pe}$ (see Fig.~\ref{fig:MAN2shear}), it is helpful to regard the breakup kernels $K_i^\text{TB}$, $K_i^\text{SB}$, and $K_{ij}^\text{SCB}$. The frequency of shear-induced breakup of a single aggregate increases linearly with increasing P{\'e}clet number $\text{Pe}$, and it should surpass the rate of thermal breakup when $K_i^\text{SB}/K_i^\text{TB} \geq 1$. Comparing Eqs.~\eqref{eq:KTB} and \eqref{eq:KSB} then yields
\begin{equation}
    \text{Pe} \geq \frac{d F_\text{bind}}{k_\text{B}T M_i^{(1/3+\alpha)}} .
    \label{eq:KTBoverKSB}
\end{equation}
For the parameters used in this work, {\it i.e.}, $d=5\,a_\text{v}$, $F_\text{bind} = 108\,\varepsilon/a_\text{v}$ and $k_\text{B}T \sim \varepsilon$, the shear forces exerted on a single rod-shaped aggregate with $M_i = 4$ become comparable to the thermal forces at $\text{Pe} \geq 85$, while the threshold decreases to $\text{Pe} \geq 10$ for a longer rod with $M_i = 20$.

If the suspension is not infinitely dilute, then collisions between clusters can also lead to the breakup of clusters. Such cooperative effects should become relevant when $K_{ij}^\text{SCB}/K_{ij}^\text{SCA} \geq 1$. Combining Eqs.~\eqref{eq:KSCA} and \eqref{eq:KSCB} leads to
\begin{equation}
    \left(W_{ij}^{-1} - 1\right) P_\text{SCB} \geq 1 ,
    \label{eq:KSCBoverKSCA}
\end{equation}
which holds as soon as $P_\text{SCB} \approx 1$, since $\left(W_{ij}^{-1} - 1\right) > 1$ for all investigated cases. This condition is fulfilled when the kinetic energy of impact $U_\text{coll} \propto \text{Pe}^2$ equals or exceeds the typical binding energy between aggregated cubes, $U_\text{bind}$ (see Sec.~\ref{sec:model.KMC}). From Fig.~\ref{fig:twoPatchKernels} we see that the overall contribution of $K_{ij}^\text{SCB}$ is negligible for small P{\'e}clet numbers, but then suddenly jumps up near $\text{Pe} \sim 10^{-1}$ and surpasses all other rate kernels for larger $\text{Pe}$. For collisions between larger clusters, this transition shifts to smaller P{\'e}clet numbers, since $U_\text{coll}$ scales roughly as $U_\text{coll} \propto (L_i + L_j)^2$, while $U_\text{bind}$ is independent of cluster size. Figure~\ref{fig:twoPatchKernels} also confirms the $\text{Pe}$-independent ratio $K_{ij}^\text{SCB}/K_{ij}^\text{SCA} > 1$ at sufficiently high P{\'e}clet numbers, as expected from Eq.~\eqref{eq:KSCBoverKSCA}. The importance of collective collisions for the shear-induced breakup of clusters is consistent with our MD simulations, where we observed a distinct decrease of $\la M \ra$ already at $\text{Pe} = 1$ in sheared suspensions with $\phi=0.112$ [see Fig.~\ref{fig:MAN2shear}(b)], whereas ultradilute suspensions of rod-shaped clusters ($M_i =4$) remained stable under shear at $\text{Pe} = 27$ but broke up at $\text{Pe} = 198$.

\section{Conclusions}
We employed a combination of molecular dynamics (MD) and rejection-free Kinetic Monte Carlo (KMC) simulations to investigate the self-assembly of amphiphilic nanocubes, which can be regarded as coarse-grained models for folded proteins\cite{liu:jcp:2007, mcmanus:coc:2016, du:nl:2021} or DNA-functionalized nanoparticles.\cite{zhang:jcp:2023} We simulated suspensions under shear and at rest, focusing on volume fractions below the freezing transitions of hard cubes. We systematically varied the number and arrangement of hydrophobic faces on the cubes, and always included a fraction of one-patch cubes to limit growth into finite-sized aggregates. This strategy allowed the assembly into superstructures that are unattainable with spherical particles, such as elongated rods and fractal objects, depending on the number and placement of hydrophobic faces. For weak hydrophobic interactions (or high temperatures), the nanocubes continuously aggregated and dissociated from each other, resulting in small aggregates of only few nanocubes. For more hydrophobic interactions (or lower temperatures), the aggregation essentially became irreversible, so that the clusters grew continuously until all hydrophobic surfaces were covered by one-patch cubes. Mixtures containing three-patch cubes exhibited extremely slow self-assembly dynamics caused by the gradual recombination of the aggregates, highlighting the challenge of equilibrating such systems using conventional MD simulations. In all investigated cases, we found that the final aggregate size increased distinctly as the fraction of multi-patch cubes in the mixtures was increased. 

In shear flow, the advective motion of the nanocubes led to faster aggregation compared to the diffusion-driven self-assembly at rest, but the steady-state average cluster size decreased with increasing shear rate (or P{\'e}clet number). We rationalize this behavior by comparing the magnitudes of the relevant rate kernels in the KMC simulations, finding that the shear-induced breakup due to collisions between aggregates becomes the dominant event at high P{\'e}clet numbers. The results from MD and KMC were in excellent quantitative agreement for the equilibrium simulations, and in semi-quantitative agreement for the shear simulations. We attribute the (minor) differences between the two approaches primarily to the smaller number of samples in the MD simulations and the approximations made in the KMC model. This complementary approach combines KMC for rapid parameter space exploration with MD for detailed analysis. It can be extended to study various particle compositions, shapes, and sizes, facilitating the rational design of building blocks for self-assembly into hierarchical superstructures.

\section*{Acknowledgments}
This work was supported by the Deutsche Forschungsgemeinschaft (DFG, German Research Foundation) through Project Nos. 233630050, 405552959, and 470113688. Y.K. was supported by JSPS KAKENHI Grant Number JP21K20411. This research was supported in part by the National Science Foundation under Grant No. NSF PHY-1748958.

\bibliography{references}

\begin{thebibliography}{56}%
\makeatletter
\providecommand \@ifxundefined [1]{%
 \@ifx{#1\undefined}
}%
\providecommand \@ifnum [1]{%
 \ifnum #1\expandafter \@firstoftwo
 \else \expandafter \@secondoftwo
 \fi
}%
\providecommand \@ifx [1]{%
 \ifx #1\expandafter \@firstoftwo
 \else \expandafter \@secondoftwo
 \fi
}%
\providecommand \natexlab [1]{#1}%
\providecommand \enquote  [1]{``#1''}%
\providecommand \bibnamefont  [1]{#1}%
\providecommand \bibfnamefont [1]{#1}%
\providecommand \citenamefont [1]{#1}%
\providecommand \href@noop [0]{\@secondoftwo}%
\providecommand \href [0]{\begingroup \@sanitize@url \@href}%
\providecommand \@href[1]{\@@startlink{#1}\@@href}%
\providecommand \@@href[1]{\endgroup#1\@@endlink}%
\providecommand \@sanitize@url [0]{\catcode `\\12\catcode `\$12\catcode
  `\&12\catcode `\#12\catcode `\^12\catcode `\_12\catcode `\%12\relax}%
\providecommand \@@startlink[1]{}%
\providecommand \@@endlink[0]{}%
\providecommand \url  [0]{\begingroup\@sanitize@url \@url }%
\providecommand \@url [1]{\endgroup\@href {#1}{\urlprefix }}%
\providecommand \urlprefix  [0]{URL }%
\providecommand \Eprint [0]{\href }%
\providecommand \doibase [0]{http://dx.doi.org/}%
\providecommand \selectlanguage [0]{\@gobble}%
\providecommand \bibinfo  [0]{\@secondoftwo}%
\providecommand \bibfield  [0]{\@secondoftwo}%
\providecommand \translation [1]{[#1]}%
\providecommand \BibitemOpen [0]{}%
\providecommand \bibitemStop [0]{}%
\providecommand \bibitemNoStop [0]{.\EOS\space}%
\providecommand \EOS [0]{\spacefactor3000\relax}%
\providecommand \BibitemShut  [1]{\csname bibitem#1\endcsname}%
\let\auto@bib@innerbib\@empty
\bibitem [{\citenamefont {Antonietti}\ and\ \citenamefont
  {G{\"o}ltner}(1997)}]{antonietti:ac:1997}%
  \BibitemOpen
  \bibfield  {author} {\bibinfo {author} {\bibfnamefont {M.}~\bibnamefont
  {Antonietti}}\ and\ \bibinfo {author} {\bibfnamefont {C.}~\bibnamefont
  {G{\"o}ltner}},\ }\href@noop {} {\bibfield  {journal} {\bibinfo  {journal}
  {Angew. Chem., Int. Ed.}\ }\textbf {\bibinfo {volume} {36}},\ \bibinfo
  {pages} {910} (\bibinfo {year} {1997})}\BibitemShut {NoStop}%
\bibitem [{\citenamefont {Pawar}\ and\ \citenamefont
  {Kretzschmar}(2010)}]{pawar:mrc:2010}%
  \BibitemOpen
  \bibfield  {author} {\bibinfo {author} {\bibfnamefont {A.~B.}\ \bibnamefont
  {Pawar}}\ and\ \bibinfo {author} {\bibfnamefont {I.}~\bibnamefont
  {Kretzschmar}},\ }\href@noop {} {\bibfield  {journal} {\bibinfo  {journal}
  {Macromol. Rapid Commun.}\ }\textbf {\bibinfo {volume} {31}},\ \bibinfo
  {pages} {150} (\bibinfo {year} {2010})}\BibitemShut {NoStop}%
\bibitem [{\citenamefont {Glotzer}\ and\ \citenamefont
  {Solomon}(2007)}]{glotzer:nm:2007}%
  \BibitemOpen
  \bibfield  {author} {\bibinfo {author} {\bibfnamefont {S.~C.}\ \bibnamefont
  {Glotzer}}\ and\ \bibinfo {author} {\bibfnamefont {M.~J.}\ \bibnamefont
  {Solomon}},\ }\href@noop {} {\bibfield  {journal} {\bibinfo  {journal} {Nat.
  Mater.}\ }\textbf {\bibinfo {volume} {6}},\ \bibinfo {pages} {557} (\bibinfo
  {year} {2007})}\BibitemShut {NoStop}%
\bibitem [{\citenamefont {Li}\ \emph {et~al.}(2020)\citenamefont {Li},
  \citenamefont {Palis}, \citenamefont {M{\'e}rindol}, \citenamefont {Majimel},
  \citenamefont {Ravaine},\ and\ \citenamefont {Duguet}}]{li:csr:2020}%
  \BibitemOpen
  \bibfield  {author} {\bibinfo {author} {\bibfnamefont {W.}~\bibnamefont
  {Li}}, \bibinfo {author} {\bibfnamefont {H.}~\bibnamefont {Palis}}, \bibinfo
  {author} {\bibfnamefont {R.}~\bibnamefont {M{\'e}rindol}}, \bibinfo {author}
  {\bibfnamefont {J.}~\bibnamefont {Majimel}}, \bibinfo {author} {\bibfnamefont
  {S.}~\bibnamefont {Ravaine}}, \ and\ \bibinfo {author} {\bibfnamefont
  {E.}~\bibnamefont {Duguet}},\ }\href@noop {} {\bibfield  {journal} {\bibinfo
  {journal} {Chem. Soc. Rev.}\ }\textbf {\bibinfo {volume} {49}},\ \bibinfo
  {pages} {1955} (\bibinfo {year} {2020})}\BibitemShut {NoStop}%
\bibitem [{\citenamefont {Sciortino}, \citenamefont {Giacometti},\ and\
  \citenamefont {Pastore}(2009)}]{sciortino:prl:2009}%
  \BibitemOpen
  \bibfield  {author} {\bibinfo {author} {\bibfnamefont {F.}~\bibnamefont
  {Sciortino}}, \bibinfo {author} {\bibfnamefont {A.}~\bibnamefont
  {Giacometti}}, \ and\ \bibinfo {author} {\bibfnamefont {G.}~\bibnamefont
  {Pastore}},\ }\href@noop {} {\bibfield  {journal} {\bibinfo  {journal} {Phys.
  Rev. Lett.}\ }\textbf {\bibinfo {volume} {103}},\ \bibinfo {pages} {237801}
  (\bibinfo {year} {2009})}\BibitemShut {NoStop}%
\bibitem [{\citenamefont {Bianchi}, \citenamefont {Panagiotopoulos},\ and\
  \citenamefont {Nikoubashman}(2015)}]{bianchi:sm:2015}%
  \BibitemOpen
  \bibfield  {author} {\bibinfo {author} {\bibfnamefont {E.}~\bibnamefont
  {Bianchi}}, \bibinfo {author} {\bibfnamefont {A.~Z.}\ \bibnamefont
  {Panagiotopoulos}}, \ and\ \bibinfo {author} {\bibfnamefont {A.}~\bibnamefont
  {Nikoubashman}},\ }\href@noop {} {\bibfield  {journal} {\bibinfo  {journal}
  {Soft Matter}\ }\textbf {\bibinfo {volume} {11}},\ \bibinfo {pages} {3767}
  (\bibinfo {year} {2015})}\BibitemShut {NoStop}%
\bibitem [{\citenamefont {Kobayashi}\ and\ \citenamefont
  {Arai}(2016)}]{kobayashi:sm:2016}%
  \BibitemOpen
  \bibfield  {author} {\bibinfo {author} {\bibfnamefont {Y.}~\bibnamefont
  {Kobayashi}}\ and\ \bibinfo {author} {\bibfnamefont {N.}~\bibnamefont
  {Arai}},\ }\href@noop {} {\bibfield  {journal} {\bibinfo  {journal} {Soft
  Matter}\ }\textbf {\bibinfo {volume} {12}},\ \bibinfo {pages} {378} (\bibinfo
  {year} {2016})}\BibitemShut {NoStop}%
\bibitem [{\citenamefont {Kobayashi}, \citenamefont {Arai},\ and\ \citenamefont
  {Nikoubashman}(2020{\natexlab{a}})}]{kobayashi:sm:2020}%
  \BibitemOpen
  \bibfield  {author} {\bibinfo {author} {\bibfnamefont {Y.}~\bibnamefont
  {Kobayashi}}, \bibinfo {author} {\bibfnamefont {N.}~\bibnamefont {Arai}}, \
  and\ \bibinfo {author} {\bibfnamefont {A.}~\bibnamefont {Nikoubashman}},\
  }\href@noop {} {\bibfield  {journal} {\bibinfo  {journal} {Soft Matter}\
  }\textbf {\bibinfo {volume} {16}},\ \bibinfo {pages} {476} (\bibinfo {year}
  {2020}{\natexlab{a}})}\BibitemShut {NoStop}%
\bibitem [{\citenamefont {Bianchi}\ \emph {et~al.}(2006)\citenamefont
  {Bianchi}, \citenamefont {Largo}, \citenamefont {Tartaglia}, \citenamefont
  {Zaccarelli},\ and\ \citenamefont {Sciortino}}]{bianchi:prl:2006}%
  \BibitemOpen
  \bibfield  {author} {\bibinfo {author} {\bibfnamefont {E.}~\bibnamefont
  {Bianchi}}, \bibinfo {author} {\bibfnamefont {J.}~\bibnamefont {Largo}},
  \bibinfo {author} {\bibfnamefont {P.}~\bibnamefont {Tartaglia}}, \bibinfo
  {author} {\bibfnamefont {E.}~\bibnamefont {Zaccarelli}}, \ and\ \bibinfo
  {author} {\bibfnamefont {F.}~\bibnamefont {Sciortino}},\ }\href@noop {}
  {\bibfield  {journal} {\bibinfo  {journal} {Phys. Rev. Lett.}\ }\textbf
  {\bibinfo {volume} {97}},\ \bibinfo {pages} {168301} (\bibinfo {year}
  {2006})}\BibitemShut {NoStop}%
\bibitem [{\citenamefont {Sciortino}\ and\ \citenamefont
  {Zaccarelli}(2017)}]{sciortino:cocis:2017}%
  \BibitemOpen
  \bibfield  {author} {\bibinfo {author} {\bibfnamefont {F.}~\bibnamefont
  {Sciortino}}\ and\ \bibinfo {author} {\bibfnamefont {E.}~\bibnamefont
  {Zaccarelli}},\ }\href@noop {} {\bibfield  {journal} {\bibinfo  {journal}
  {Curr. Opin. Colloid Interface Sci.}\ }\textbf {\bibinfo {volume} {30}},\
  \bibinfo {pages} {90} (\bibinfo {year} {2017})}\BibitemShut {NoStop}%
\bibitem [{\citenamefont {Tran}, \citenamefont {Lesieur},\ and\ \citenamefont
  {Faivre}(2014)}]{tran:eodd:2014}%
  \BibitemOpen
  \bibfield  {author} {\bibinfo {author} {\bibfnamefont {L.-T.-C.}\
  \bibnamefont {Tran}}, \bibinfo {author} {\bibfnamefont {S.}~\bibnamefont
  {Lesieur}}, \ and\ \bibinfo {author} {\bibfnamefont {V.}~\bibnamefont
  {Faivre}},\ }\href@noop {} {\bibfield  {journal} {\bibinfo  {journal} {Expert
  Opin. Drug Deliv.}\ }\textbf {\bibinfo {volume} {11}},\ \bibinfo {pages}
  {1061} (\bibinfo {year} {2014})}\BibitemShut {NoStop}%
\bibitem [{\citenamefont {Su}\ \emph {et~al.}(2019)\citenamefont {Su},
  \citenamefont {Hurd~Price}, \citenamefont {Jing}, \citenamefont {Tian},
  \citenamefont {Liu},\ and\ \citenamefont {Qian}}]{su:mtb:2019}%
  \BibitemOpen
  \bibfield  {author} {\bibinfo {author} {\bibfnamefont {H.}~\bibnamefont
  {Su}}, \bibinfo {author} {\bibfnamefont {C.-A.}\ \bibnamefont {Hurd~Price}},
  \bibinfo {author} {\bibfnamefont {L.}~\bibnamefont {Jing}}, \bibinfo {author}
  {\bibfnamefont {Q.}~\bibnamefont {Tian}}, \bibinfo {author} {\bibfnamefont
  {J.}~\bibnamefont {Liu}}, \ and\ \bibinfo {author} {\bibfnamefont
  {K.}~\bibnamefont {Qian}},\ }\href@noop {} {\bibfield  {journal} {\bibinfo
  {journal} {Mater. Today Bio}\ }\textbf {\bibinfo {volume} {4}},\ \bibinfo
  {pages} {100033} (\bibinfo {year} {2019})}\BibitemShut {NoStop}%
\bibitem [{\citenamefont {Bradley}\ \emph {et~al.}(2017)\citenamefont
  {Bradley}, \citenamefont {Chen}, \citenamefont {Stebe},\ and\ \citenamefont
  {Lee}}]{bradley:cocis:2017}%
  \BibitemOpen
  \bibfield  {author} {\bibinfo {author} {\bibfnamefont {L.~C.}\ \bibnamefont
  {Bradley}}, \bibinfo {author} {\bibfnamefont {W.-H.}\ \bibnamefont {Chen}},
  \bibinfo {author} {\bibfnamefont {K.~J.}\ \bibnamefont {Stebe}}, \ and\
  \bibinfo {author} {\bibfnamefont {D.}~\bibnamefont {Lee}},\ }\href@noop {}
  {\bibfield  {journal} {\bibinfo  {journal} {Curr. Opin. Colloid Interface
  Sci.}\ }\textbf {\bibinfo {volume} {30}},\ \bibinfo {pages} {25} (\bibinfo
  {year} {2017})}\BibitemShut {NoStop}%
\bibitem [{\citenamefont {Morozova}\ and\ \citenamefont
  {Nikoubashman}(2019)}]{morozova:lng:2019}%
  \BibitemOpen
  \bibfield  {author} {\bibinfo {author} {\bibfnamefont {T.~I.}\ \bibnamefont
  {Morozova}}\ and\ \bibinfo {author} {\bibfnamefont {A.}~\bibnamefont
  {Nikoubashman}},\ }\href@noop {} {\bibfield  {journal} {\bibinfo  {journal}
  {Langmuir}\ }\textbf {\bibinfo {volume} {35}},\ \bibinfo {pages} {16907}
  (\bibinfo {year} {2019})}\BibitemShut {NoStop}%
\bibitem [{\citenamefont {Morozova}\ \emph {et~al.}(2020)\citenamefont
  {Morozova}, \citenamefont {Lee}, \citenamefont {Bizmark}, \citenamefont
  {Datta}, \citenamefont {Prud'homme}, \citenamefont {Nikoubashman},\ and\
  \citenamefont {Priestley}}]{morozova:acscs:2020}%
  \BibitemOpen
  \bibfield  {author} {\bibinfo {author} {\bibfnamefont {T.~I.}\ \bibnamefont
  {Morozova}}, \bibinfo {author} {\bibfnamefont {V.~E.}\ \bibnamefont {Lee}},
  \bibinfo {author} {\bibfnamefont {N.}~\bibnamefont {Bizmark}}, \bibinfo
  {author} {\bibfnamefont {S.~S.}\ \bibnamefont {Datta}}, \bibinfo {author}
  {\bibfnamefont {R.~K.}\ \bibnamefont {Prud'homme}}, \bibinfo {author}
  {\bibfnamefont {A.}~\bibnamefont {Nikoubashman}}, \ and\ \bibinfo {author}
  {\bibfnamefont {R.~D.}\ \bibnamefont {Priestley}},\ }\href@noop {} {\bibfield
   {journal} {\bibinfo  {journal} {ACS Cent. Sci.}\ }\textbf {\bibinfo {volume}
  {6}},\ \bibinfo {pages} {166} (\bibinfo {year} {2020})}\BibitemShut {NoStop}%
\bibitem [{\citenamefont {Correia}, \citenamefont {Brown},\ and\ \citenamefont
  {Razavi}(2021)}]{correia:nm:2021}%
  \BibitemOpen
  \bibfield  {author} {\bibinfo {author} {\bibfnamefont {E.~L.}\ \bibnamefont
  {Correia}}, \bibinfo {author} {\bibfnamefont {N.}~\bibnamefont {Brown}}, \
  and\ \bibinfo {author} {\bibfnamefont {S.}~\bibnamefont {Razavi}},\
  }\href@noop {} {\bibfield  {journal} {\bibinfo  {journal} {Nanomater.}\
  }\textbf {\bibinfo {volume} {11}},\ \bibinfo {pages} {374} (\bibinfo {year}
  {2021})}\BibitemShut {NoStop}%
\bibitem [{\citenamefont {Kirillova}\ \emph {et~al.}(2015)\citenamefont
  {Kirillova}, \citenamefont {Schliebe}, \citenamefont {Stoychev},
  \citenamefont {Jakob}, \citenamefont {Lang},\ and\ \citenamefont
  {Synytska}}]{kirillova:acsami:2015}%
  \BibitemOpen
  \bibfield  {author} {\bibinfo {author} {\bibfnamefont {A.}~\bibnamefont
  {Kirillova}}, \bibinfo {author} {\bibfnamefont {C.}~\bibnamefont {Schliebe}},
  \bibinfo {author} {\bibfnamefont {G.}~\bibnamefont {Stoychev}}, \bibinfo
  {author} {\bibfnamefont {A.}~\bibnamefont {Jakob}}, \bibinfo {author}
  {\bibfnamefont {H.}~\bibnamefont {Lang}}, \ and\ \bibinfo {author}
  {\bibfnamefont {A.}~\bibnamefont {Synytska}},\ }\href@noop {} {\bibfield
  {journal} {\bibinfo  {journal} {ACS Appl. Mater. Interfaces}\ }\textbf
  {\bibinfo {volume} {7}},\ \bibinfo {pages} {21218} (\bibinfo {year}
  {2015})}\BibitemShut {NoStop}%
\bibitem [{\citenamefont {Marschelke}, \citenamefont {Fery},\ and\
  \citenamefont {Synytska}(2020)}]{marschelke:cps:2020}%
  \BibitemOpen
  \bibfield  {author} {\bibinfo {author} {\bibfnamefont {C.}~\bibnamefont
  {Marschelke}}, \bibinfo {author} {\bibfnamefont {A.}~\bibnamefont {Fery}}, \
  and\ \bibinfo {author} {\bibfnamefont {A.}~\bibnamefont {Synytska}},\
  }\href@noop {} {\bibfield  {journal} {\bibinfo  {journal} {Colloid Polym.
  Sci}\ }\textbf {\bibinfo {volume} {298}},\ \bibinfo {pages} {841} (\bibinfo
  {year} {2020})}\BibitemShut {NoStop}%
\bibitem [{\citenamefont {Beijerinck}(1898)}]{beijerinck:vka:1898}%
  \BibitemOpen
  \bibfield  {author} {\bibinfo {author} {\bibfnamefont {M.~W.}\ \bibnamefont
  {Beijerinck}},\ }\href@noop {} {\bibfield  {journal} {\bibinfo  {journal}
  {Verh. K. Akad. Wet. Amsterdam, Afd. Natuurkd.}\ }\textbf {\bibinfo {volume}
  {5}},\ \bibinfo {pages} {3} (\bibinfo {year} {1898})}\BibitemShut {NoStop}%
\bibitem [{\citenamefont {Bawden}\ \emph {et~al.}(1936)\citenamefont {Bawden},
  \citenamefont {Pirie}, \citenamefont {Bernal},\ and\ \citenamefont
  {Fankuchen}}]{bawden:nat:1936}%
  \BibitemOpen
  \bibfield  {author} {\bibinfo {author} {\bibfnamefont {F.~C.}\ \bibnamefont
  {Bawden}}, \bibinfo {author} {\bibfnamefont {N.~W.}\ \bibnamefont {Pirie}},
  \bibinfo {author} {\bibfnamefont {J.~D.}\ \bibnamefont {Bernal}}, \ and\
  \bibinfo {author} {\bibfnamefont {I.}~\bibnamefont {Fankuchen}},\ }\href@noop
  {} {\bibfield  {journal} {\bibinfo  {journal} {Nature}\ }\textbf {\bibinfo
  {volume} {138}},\ \bibinfo {pages} {1051} (\bibinfo {year}
  {1936})}\BibitemShut {NoStop}%
\bibitem [{\citenamefont {van~der Kooij}, \citenamefont {Kassapidou},\ and\
  \citenamefont {Lekkerkerker}(2000)}]{kooij:nat:2000}%
  \BibitemOpen
  \bibfield  {author} {\bibinfo {author} {\bibfnamefont {F.~M.}\ \bibnamefont
  {van~der Kooij}}, \bibinfo {author} {\bibfnamefont {K.}~\bibnamefont
  {Kassapidou}}, \ and\ \bibinfo {author} {\bibfnamefont {H.~N.~W.}\
  \bibnamefont {Lekkerkerker}},\ }\href@noop {} {\bibfield  {journal} {\bibinfo
   {journal} {Nature}\ }\textbf {\bibinfo {volume} {406}},\ \bibinfo {pages}
  {868} (\bibinfo {year} {2000})}\BibitemShut {NoStop}%
\bibitem [{\citenamefont {John}, \citenamefont {Stroock},\ and\ \citenamefont
  {Escobedo}(2004)}]{john:jcp:2004}%
  \BibitemOpen
  \bibfield  {author} {\bibinfo {author} {\bibfnamefont {B.~S.}\ \bibnamefont
  {John}}, \bibinfo {author} {\bibfnamefont {A.}~\bibnamefont {Stroock}}, \
  and\ \bibinfo {author} {\bibfnamefont {F.~A.}\ \bibnamefont {Escobedo}},\
  }\href@noop {} {\bibfield  {journal} {\bibinfo  {journal} {J. Chem. Phys.}\
  }\textbf {\bibinfo {volume} {120}},\ \bibinfo {pages} {9383} (\bibinfo {year}
  {2004})}\BibitemShut {NoStop}%
\bibitem [{\citenamefont {Agarwal}\ and\ \citenamefont
  {Escobedo}(2011)}]{agarwal:nm:2011}%
  \BibitemOpen
  \bibfield  {author} {\bibinfo {author} {\bibfnamefont {U.}~\bibnamefont
  {Agarwal}}\ and\ \bibinfo {author} {\bibfnamefont {F.~A.}\ \bibnamefont
  {Escobedo}},\ }\href@noop {} {\bibfield  {journal} {\bibinfo  {journal} {Nat.
  Mater.}\ }\textbf {\bibinfo {volume} {10}},\ \bibinfo {pages} {230} (\bibinfo
  {year} {2011})}\BibitemShut {NoStop}%
\bibitem [{\citenamefont {Damasceno}, \citenamefont {Engel},\ and\
  \citenamefont {Glotzer}(2012)}]{damasceno:sci:2012}%
  \BibitemOpen
  \bibfield  {author} {\bibinfo {author} {\bibfnamefont {P.~F.}\ \bibnamefont
  {Damasceno}}, \bibinfo {author} {\bibfnamefont {M.}~\bibnamefont {Engel}}, \
  and\ \bibinfo {author} {\bibfnamefont {S.~C.}\ \bibnamefont {Glotzer}},\
  }\href@noop {} {\bibfield  {journal} {\bibinfo  {journal} {Science}\ }\textbf
  {\bibinfo {volume} {337}},\ \bibinfo {pages} {453} (\bibinfo {year}
  {2012})}\BibitemShut {NoStop}%
\bibitem [{\citenamefont {Smallenburg}\ \emph {et~al.}(2012)\citenamefont
  {Smallenburg}, \citenamefont {Filion}, \citenamefont {Marechal},\ and\
  \citenamefont {Dijkstra}}]{smallenburg:pnas:2012}%
  \BibitemOpen
  \bibfield  {author} {\bibinfo {author} {\bibfnamefont {F.}~\bibnamefont
  {Smallenburg}}, \bibinfo {author} {\bibfnamefont {L.}~\bibnamefont {Filion}},
  \bibinfo {author} {\bibfnamefont {M.}~\bibnamefont {Marechal}}, \ and\
  \bibinfo {author} {\bibfnamefont {M.}~\bibnamefont {Dijkstra}},\ }\href@noop
  {} {\bibfield  {journal} {\bibinfo  {journal} {Proc. Natl. Acad. Sci.
  U.S.A.}\ }\textbf {\bibinfo {volume} {109}},\ \bibinfo {pages} {17886}
  (\bibinfo {year} {2012})}\BibitemShut {NoStop}%
\bibitem [{\citenamefont {Wang}, \citenamefont {Min},\ and\ \citenamefont
  {Yu}(2007)}]{wang:jpcc:2007}%
  \BibitemOpen
  \bibfield  {author} {\bibinfo {author} {\bibfnamefont {S.-B.}\ \bibnamefont
  {Wang}}, \bibinfo {author} {\bibfnamefont {Y.-L.}\ \bibnamefont {Min}}, \
  and\ \bibinfo {author} {\bibfnamefont {S.-H.}\ \bibnamefont {Yu}},\
  }\href@noop {} {\bibfield  {journal} {\bibinfo  {journal} {J. Phys. Chem. C}\
  }\textbf {\bibinfo {volume} {111}},\ \bibinfo {pages} {3551} (\bibinfo {year}
  {2007})}\BibitemShut {NoStop}%
\bibitem [{\citenamefont {Mayer}\ \emph {et~al.}(2017)\citenamefont {Mayer},
  \citenamefont {Steiner}, \citenamefont {R{\"o}der}, \citenamefont {Formanek},
  \citenamefont {K{\"o}nig},\ and\ \citenamefont {Fery}}]{mayer:ac:2017}%
  \BibitemOpen
  \bibfield  {author} {\bibinfo {author} {\bibfnamefont {M.}~\bibnamefont
  {Mayer}}, \bibinfo {author} {\bibfnamefont {A.~M.}\ \bibnamefont {Steiner}},
  \bibinfo {author} {\bibfnamefont {F.}~\bibnamefont {R{\"o}der}}, \bibinfo
  {author} {\bibfnamefont {P.}~\bibnamefont {Formanek}}, \bibinfo {author}
  {\bibfnamefont {T.~A.~F.}\ \bibnamefont {K{\"o}nig}}, \ and\ \bibinfo
  {author} {\bibfnamefont {A.}~\bibnamefont {Fery}},\ }\href@noop {} {\bibfield
   {journal} {\bibinfo  {journal} {Angew. Chem., Int. Ed.}\ }\textbf {\bibinfo
  {volume} {56}},\ \bibinfo {pages} {15866} (\bibinfo {year}
  {2017})}\BibitemShut {NoStop}%
\bibitem [{\citenamefont {Toso}, \citenamefont {Baranov},\ and\ \citenamefont
  {Manna}(2021)}]{toso:acs:2021}%
  \BibitemOpen
  \bibfield  {author} {\bibinfo {author} {\bibfnamefont {S.}~\bibnamefont
  {Toso}}, \bibinfo {author} {\bibfnamefont {D.}~\bibnamefont {Baranov}}, \
  and\ \bibinfo {author} {\bibfnamefont {L.}~\bibnamefont {Manna}},\
  }\href@noop {} {\bibfield  {journal} {\bibinfo  {journal} {Acc. Chem. Res.}\
  }\textbf {\bibinfo {volume} {54}},\ \bibinfo {pages} {498} (\bibinfo {year}
  {2021})}\BibitemShut {NoStop}%
\bibitem [{\citenamefont {Kobayashi}\ and\ \citenamefont
  {Nikoubashman}(2022)}]{kobayashi:lng:2022}%
  \BibitemOpen
  \bibfield  {author} {\bibinfo {author} {\bibfnamefont {Y.}~\bibnamefont
  {Kobayashi}}\ and\ \bibinfo {author} {\bibfnamefont {A.}~\bibnamefont
  {Nikoubashman}},\ }\href@noop {} {\bibfield  {journal} {\bibinfo  {journal}
  {Langmuir}\ }\textbf {\bibinfo {volume} {38}},\ \bibinfo {pages} {10642}
  (\bibinfo {year} {2022})}\BibitemShut {NoStop}%
\bibitem [{\citenamefont {Zhang}\ \emph {et~al.}(2023)\citenamefont {Zhang},
  \citenamefont {Giunta}, \citenamefont {Liang},\ and\ \citenamefont
  {Dijkstra}}]{zhang:jcp:2023}%
  \BibitemOpen
  \bibfield  {author} {\bibinfo {author} {\bibfnamefont {Y.}~\bibnamefont
  {Zhang}}, \bibinfo {author} {\bibfnamefont {G.}~\bibnamefont {Giunta}},
  \bibinfo {author} {\bibfnamefont {H.}~\bibnamefont {Liang}}, \ and\ \bibinfo
  {author} {\bibfnamefont {M.}~\bibnamefont {Dijkstra}},\ }\href@noop {}
  {\bibfield  {journal} {\bibinfo  {journal} {J. Chem. Phys.}\ }\textbf
  {\bibinfo {volume} {158}},\ \bibinfo {pages} {184902} (\bibinfo {year}
  {2023})}\BibitemShut {NoStop}%
\bibitem [{\citenamefont {Ru{\c s}en~Argun}\ and\ \citenamefont
  {Statt}(2023)}]{argun:arxiv:2023}%
  \BibitemOpen
  \bibfield  {author} {\bibinfo {author} {\bibfnamefont {B.}~\bibnamefont
  {Ru{\c s}en~Argun}}\ and\ \bibinfo {author} {\bibfnamefont {A.}~\bibnamefont
  {Statt}},\ }\href@noop {} {\bibfield  {journal} {\bibinfo  {journal} {arXiv}\
  ,\ \bibinfo {pages} {2305.05453}} (\bibinfo {year} {2023})}\BibitemShut
  {NoStop}%
\bibitem [{\citenamefont {Rothemund}(2006)}]{rothemund:nat:2006}%
  \BibitemOpen
  \bibfield  {author} {\bibinfo {author} {\bibfnamefont {P.~W.~K.}\
  \bibnamefont {Rothemund}},\ }\href@noop {} {\bibfield  {journal} {\bibinfo
  {journal} {Nature}\ }\textbf {\bibinfo {volume} {440}},\ \bibinfo {pages}
  {297} (\bibinfo {year} {2006})}\BibitemShut {NoStop}%
\bibitem [{\citenamefont {Seeman}\ and\ \citenamefont
  {Sleiman}(2017)}]{seeman:nr:2017}%
  \BibitemOpen
  \bibfield  {author} {\bibinfo {author} {\bibfnamefont {N.~C.}\ \bibnamefont
  {Seeman}}\ and\ \bibinfo {author} {\bibfnamefont {H.~F.}\ \bibnamefont
  {Sleiman}},\ }\href@noop {} {\bibfield  {journal} {\bibinfo  {journal} {Nat.
  Rev. Mater.}\ }\textbf {\bibinfo {volume} {3}},\ \bibinfo {pages} {17068}
  (\bibinfo {year} {2017})}\BibitemShut {NoStop}%
\bibitem [{\citenamefont {Dey}\ \emph {et~al.}(2021)\citenamefont {Dey},
  \citenamefont {Fan}, \citenamefont {Gothelf}, \citenamefont {Li},
  \citenamefont {Lin}, \citenamefont {Liu}, \citenamefont {Liu}, \citenamefont
  {Nijenhuis}, \citenamefont {Sacc{\`a}}, \citenamefont {Simmel}, \citenamefont
  {Yan},\ and\ \citenamefont {Zhan}}]{dey:nr:2021}%
  \BibitemOpen
  \bibfield  {author} {\bibinfo {author} {\bibfnamefont {S.}~\bibnamefont
  {Dey}}, \bibinfo {author} {\bibfnamefont {C.}~\bibnamefont {Fan}}, \bibinfo
  {author} {\bibfnamefont {K.~V.}\ \bibnamefont {Gothelf}}, \bibinfo {author}
  {\bibfnamefont {J.}~\bibnamefont {Li}}, \bibinfo {author} {\bibfnamefont
  {C.}~\bibnamefont {Lin}}, \bibinfo {author} {\bibfnamefont {L.}~\bibnamefont
  {Liu}}, \bibinfo {author} {\bibfnamefont {N.}~\bibnamefont {Liu}}, \bibinfo
  {author} {\bibfnamefont {M.~A.~D.}\ \bibnamefont {Nijenhuis}}, \bibinfo
  {author} {\bibfnamefont {B.}~\bibnamefont {Sacc{\`a}}}, \bibinfo {author}
  {\bibfnamefont {F.~C.}\ \bibnamefont {Simmel}}, \bibinfo {author}
  {\bibfnamefont {H.}~\bibnamefont {Yan}}, \ and\ \bibinfo {author}
  {\bibfnamefont {P.}~\bibnamefont {Zhan}},\ }\href@noop {} {\bibfield
  {journal} {\bibinfo  {journal} {Nat. Rev. Dis. Primers}\ }\textbf {\bibinfo
  {volume} {1}},\ \bibinfo {pages} {13} (\bibinfo {year} {2021})}\BibitemShut
  {NoStop}%
\bibitem [{\citenamefont {Scheible}\ \emph {et~al.}(2015)\citenamefont
  {Scheible}, \citenamefont {Ong}, \citenamefont {Woehrstein}, \citenamefont
  {Jungmann}, \citenamefont {Yin},\ and\ \citenamefont
  {Simmel}}]{scheible:small:2015}%
  \BibitemOpen
  \bibfield  {author} {\bibinfo {author} {\bibfnamefont {M.~B.}\ \bibnamefont
  {Scheible}}, \bibinfo {author} {\bibfnamefont {L.~L.}\ \bibnamefont {Ong}},
  \bibinfo {author} {\bibfnamefont {J.~B.}\ \bibnamefont {Woehrstein}},
  \bibinfo {author} {\bibfnamefont {R.}~\bibnamefont {Jungmann}}, \bibinfo
  {author} {\bibfnamefont {P.}~\bibnamefont {Yin}}, \ and\ \bibinfo {author}
  {\bibfnamefont {F.~C.}\ \bibnamefont {Simmel}},\ }\href@noop {} {\bibfield
  {journal} {\bibinfo  {journal} {Small}\ }\textbf {\bibinfo {volume} {11}},\
  \bibinfo {pages} {5200} (\bibinfo {year} {2015})}\BibitemShut {NoStop}%
\bibitem [{\citenamefont {Liu}, \citenamefont {Kumar},\ and\ \citenamefont
  {Sciortino}(2007)}]{liu:jcp:2007}%
  \BibitemOpen
  \bibfield  {author} {\bibinfo {author} {\bibfnamefont {H.}~\bibnamefont
  {Liu}}, \bibinfo {author} {\bibfnamefont {S.~K.}\ \bibnamefont {Kumar}}, \
  and\ \bibinfo {author} {\bibfnamefont {F.}~\bibnamefont {Sciortino}},\
  }\href@noop {} {\bibfield  {journal} {\bibinfo  {journal} {J. Chem. Phys.}\
  }\textbf {\bibinfo {volume} {127}},\ \bibinfo {pages} {084902} (\bibinfo
  {year} {2007})}\BibitemShut {NoStop}%
\bibitem [{\citenamefont {McManus}\ \emph {et~al.}(2016)\citenamefont
  {McManus}, \citenamefont {Charbonneau}, \citenamefont {Zaccarelli},\ and\
  \citenamefont {Asherie}}]{mcmanus:coc:2016}%
  \BibitemOpen
  \bibfield  {author} {\bibinfo {author} {\bibfnamefont {J.~J.}\ \bibnamefont
  {McManus}}, \bibinfo {author} {\bibfnamefont {P.}~\bibnamefont
  {Charbonneau}}, \bibinfo {author} {\bibfnamefont {E.}~\bibnamefont
  {Zaccarelli}}, \ and\ \bibinfo {author} {\bibfnamefont {N.}~\bibnamefont
  {Asherie}},\ }\href@noop {} {\bibfield  {journal} {\bibinfo  {journal} {Curr.
  Opin. Colloid Interface Sci.}\ }\textbf {\bibinfo {volume} {22}},\ \bibinfo
  {pages} {73} (\bibinfo {year} {2016})}\BibitemShut {NoStop}%
\bibitem [{\citenamefont {Du}\ \emph {et~al.}(2021)\citenamefont {Du},
  \citenamefont {Zhou}, \citenamefont {Yu}, \citenamefont {Zhai}, \citenamefont
  {Chen},\ and\ \citenamefont {Wang}}]{du:nl:2021}%
  \BibitemOpen
  \bibfield  {author} {\bibinfo {author} {\bibfnamefont {M.}~\bibnamefont
  {Du}}, \bibinfo {author} {\bibfnamefont {K.}~\bibnamefont {Zhou}}, \bibinfo
  {author} {\bibfnamefont {R.}~\bibnamefont {Yu}}, \bibinfo {author}
  {\bibfnamefont {Y.}~\bibnamefont {Zhai}}, \bibinfo {author} {\bibfnamefont
  {G.}~\bibnamefont {Chen}}, \ and\ \bibinfo {author} {\bibfnamefont
  {Q.}~\bibnamefont {Wang}},\ }\href@noop {} {\bibfield  {journal} {\bibinfo
  {journal} {Nano lett.}\ }\textbf {\bibinfo {volume} {21}},\ \bibinfo {pages}
  {1749} (\bibinfo {year} {2021})}\BibitemShut {NoStop}%
\bibitem [{\citenamefont {Ma}\ \emph {et~al.}(2010)\citenamefont {Ma},
  \citenamefont {Li}, \citenamefont {Cho}, \citenamefont {Li}, \citenamefont
  {Yu}, \citenamefont {Zeng}, \citenamefont {Xie},\ and\ \citenamefont
  {Xia}}]{ma:acsnano:2010}%
  \BibitemOpen
  \bibfield  {author} {\bibinfo {author} {\bibfnamefont {Y.}~\bibnamefont
  {Ma}}, \bibinfo {author} {\bibfnamefont {W.}~\bibnamefont {Li}}, \bibinfo
  {author} {\bibfnamefont {E.~C.}\ \bibnamefont {Cho}}, \bibinfo {author}
  {\bibfnamefont {Z.}~\bibnamefont {Li}}, \bibinfo {author} {\bibfnamefont
  {T.}~\bibnamefont {Yu}}, \bibinfo {author} {\bibfnamefont {J.}~\bibnamefont
  {Zeng}}, \bibinfo {author} {\bibfnamefont {Z.}~\bibnamefont {Xie}}, \ and\
  \bibinfo {author} {\bibfnamefont {Y.}~\bibnamefont {Xia}},\ }\href@noop {}
  {\bibfield  {journal} {\bibinfo  {journal} {ACS Nano}\ }\textbf {\bibinfo
  {volume} {4}},\ \bibinfo {pages} {6725} (\bibinfo {year} {2010})}\BibitemShut
  {NoStop}%
\bibitem [{\citenamefont {Royer}\ \emph {et~al.}(2015)\citenamefont {Royer},
  \citenamefont {Burton}, \citenamefont {Blair},\ and\ \citenamefont
  {Hudson}}]{royer:sm:2015}%
  \BibitemOpen
  \bibfield  {author} {\bibinfo {author} {\bibfnamefont {J.~R.}\ \bibnamefont
  {Royer}}, \bibinfo {author} {\bibfnamefont {G.~L.}\ \bibnamefont {Burton}},
  \bibinfo {author} {\bibfnamefont {D.~L.}\ \bibnamefont {Blair}}, \ and\
  \bibinfo {author} {\bibfnamefont {S.~D.}\ \bibnamefont {Hudson}},\
  }\href@noop {} {\bibfield  {journal} {\bibinfo  {journal} {Soft Matter}\
  }\textbf {\bibinfo {volume} {11}},\ \bibinfo {pages} {5656} (\bibinfo {year}
  {2015})}\BibitemShut {NoStop}%
\bibitem [{\citenamefont {Sajjadi}\ and\ \citenamefont
  {Goharshadi}(2017)}]{sajjadi:env:2017}%
  \BibitemOpen
  \bibfield  {author} {\bibinfo {author} {\bibfnamefont {S.~H.}\ \bibnamefont
  {Sajjadi}}\ and\ \bibinfo {author} {\bibfnamefont {E.~K.}\ \bibnamefont
  {Goharshadi}},\ }\href@noop {} {\bibfield  {journal} {\bibinfo  {journal} {J.
  Environ. Chem. Eng.}\ }\textbf {\bibinfo {volume} {5}},\ \bibinfo {pages}
  {1096} (\bibinfo {year} {2017})}\BibitemShut {NoStop}%
\bibitem [{\citenamefont {Humphrey}, \citenamefont {Dalke},\ and\ \citenamefont
  {Schulten}(1996)}]{vmd}%
  \BibitemOpen
  \bibfield  {author} {\bibinfo {author} {\bibfnamefont {W.}~\bibnamefont
  {Humphrey}}, \bibinfo {author} {\bibfnamefont {A.}~\bibnamefont {Dalke}}, \
  and\ \bibinfo {author} {\bibfnamefont {K.}~\bibnamefont {Schulten}},\
  }\href@noop {} {\bibfield  {journal} {\bibinfo  {journal} {J. Molec.
  Graphics}\ }\textbf {\bibinfo {volume} {14}},\ \bibinfo {pages} {33}
  (\bibinfo {year} {1996})}\BibitemShut {NoStop}%
\bibitem [{\citenamefont {Poblete}\ \emph {et~al.}(2014)\citenamefont
  {Poblete}, \citenamefont {Wysocki}, \citenamefont {Gompper},\ and\
  \citenamefont {Winkler}}]{poblete:pre:2014}%
  \BibitemOpen
  \bibfield  {author} {\bibinfo {author} {\bibfnamefont {S.}~\bibnamefont
  {Poblete}}, \bibinfo {author} {\bibfnamefont {A.}~\bibnamefont {Wysocki}},
  \bibinfo {author} {\bibfnamefont {G.}~\bibnamefont {Gompper}}, \ and\
  \bibinfo {author} {\bibfnamefont {R.~G.}\ \bibnamefont {Winkler}},\
  }\href@noop {} {\bibfield  {journal} {\bibinfo  {journal} {Phys. Rev. E}\
  }\textbf {\bibinfo {volume} {90}},\ \bibinfo {pages} {033314} (\bibinfo
  {year} {2014})}\BibitemShut {NoStop}%
\bibitem [{\citenamefont {Wani}\ \emph {et~al.}(2022)\citenamefont {Wani},
  \citenamefont {Kovakas}, \citenamefont {Nikoubashman},\ and\ \citenamefont
  {Howard}}]{wani:jcp:2022}%
  \BibitemOpen
  \bibfield  {author} {\bibinfo {author} {\bibfnamefont {Y.~M.}\ \bibnamefont
  {Wani}}, \bibinfo {author} {\bibfnamefont {P.~G.}\ \bibnamefont {Kovakas}},
  \bibinfo {author} {\bibfnamefont {A.}~\bibnamefont {Nikoubashman}}, \ and\
  \bibinfo {author} {\bibfnamefont {M.~P.}\ \bibnamefont {Howard}},\
  }\href@noop {} {\bibfield  {journal} {\bibinfo  {journal} {J. Chem. Phys.}\
  }\textbf {\bibinfo {volume} {156}},\ \bibinfo {pages} {024901} (\bibinfo
  {year} {2022})}\BibitemShut {NoStop}%
\bibitem [{\citenamefont {Weeks}, \citenamefont {Chandler},\ and\ \citenamefont
  {Andersen}(1971)}]{weeks:jcp:1971}%
  \BibitemOpen
  \bibfield  {author} {\bibinfo {author} {\bibfnamefont {J.~D.}\ \bibnamefont
  {Weeks}}, \bibinfo {author} {\bibfnamefont {D.}~\bibnamefont {Chandler}}, \
  and\ \bibinfo {author} {\bibfnamefont {H.~C.}\ \bibnamefont {Andersen}},\
  }\href@noop {} {\bibfield  {journal} {\bibinfo  {journal} {J. Chem. Phys.}\
  }\textbf {\bibinfo {volume} {54}},\ \bibinfo {pages} {5237} (\bibinfo {year}
  {1971})}\BibitemShut {NoStop}%
\bibitem [{\citenamefont {Malevanets}\ and\ \citenamefont
  {Kapral}(1999)}]{malevanets:jcp:1999}%
  \BibitemOpen
  \bibfield  {author} {\bibinfo {author} {\bibfnamefont {A.}~\bibnamefont
  {Malevanets}}\ and\ \bibinfo {author} {\bibfnamefont {R.}~\bibnamefont
  {Kapral}},\ }\href@noop {} {\bibfield  {journal} {\bibinfo  {journal} {J.
  Chem. Phys.}\ }\textbf {\bibinfo {volume} {110}},\ \bibinfo {pages} {8605}
  (\bibinfo {year} {1999})}\BibitemShut {NoStop}%
\bibitem [{\citenamefont {Gompper}\ \emph {et~al.}(2009)\citenamefont
  {Gompper}, \citenamefont {Ihle}, \citenamefont {Kroll},\ and\ \citenamefont
  {Winkler}}]{gompper:adv:2009}%
  \BibitemOpen
  \bibfield  {author} {\bibinfo {author} {\bibfnamefont {G.}~\bibnamefont
  {Gompper}}, \bibinfo {author} {\bibfnamefont {T.}~\bibnamefont {Ihle}},
  \bibinfo {author} {\bibfnamefont {D.}~\bibnamefont {Kroll}}, \ and\ \bibinfo
  {author} {\bibfnamefont {R.~G.}\ \bibnamefont {Winkler}},\ }\href@noop {}
  {\bibfield  {journal} {\bibinfo  {journal} {Adv. Polym. Sci.}\ }\textbf
  {\bibinfo {volume} {221}},\ \bibinfo {pages} {1} (\bibinfo {year}
  {2009})}\BibitemShut {NoStop}%
\bibitem [{\citenamefont {Howard}, \citenamefont {Nikoubashman},\ and\
  \citenamefont {Palmer}(2019)}]{howard:coce:2019}%
  \BibitemOpen
  \bibfield  {author} {\bibinfo {author} {\bibfnamefont {M.~P.}\ \bibnamefont
  {Howard}}, \bibinfo {author} {\bibfnamefont {A.}~\bibnamefont
  {Nikoubashman}}, \ and\ \bibinfo {author} {\bibfnamefont {J.~C.}\
  \bibnamefont {Palmer}},\ }\href@noop {} {\bibfield  {journal} {\bibinfo
  {journal} {Curr. Opin. Chem. Eng.}\ }\textbf {\bibinfo {volume} {23}},\
  \bibinfo {pages} {34} (\bibinfo {year} {2019})}\BibitemShut {NoStop}%
\bibitem [{\citenamefont {Kobayashi}, \citenamefont {Arai},\ and\ \citenamefont
  {Nikoubashman}(2020{\natexlab{b}})}]{kobayashi:lng:2020}%
  \BibitemOpen
  \bibfield  {author} {\bibinfo {author} {\bibfnamefont {Y.}~\bibnamefont
  {Kobayashi}}, \bibinfo {author} {\bibfnamefont {N.}~\bibnamefont {Arai}}, \
  and\ \bibinfo {author} {\bibfnamefont {A.}~\bibnamefont {Nikoubashman}},\
  }\href@noop {} {\bibfield  {journal} {\bibinfo  {journal} {Langmuir}\
  }\textbf {\bibinfo {volume} {36}},\ \bibinfo {pages} {14214} (\bibinfo {year}
  {2020}{\natexlab{b}})}\BibitemShut {NoStop}%
\bibitem [{\citenamefont {M{\"u}ller-Plathe}(1999)}]{mueller-plathe:pre:1999}%
  \BibitemOpen
  \bibfield  {author} {\bibinfo {author} {\bibfnamefont {F.}~\bibnamefont
  {M{\"u}ller-Plathe}},\ }\href@noop {} {\bibfield  {journal} {\bibinfo
  {journal} {Phys. Rev. E.}\ }\textbf {\bibinfo {volume} {59}},\ \bibinfo
  {pages} {4894} (\bibinfo {year} {1999})}\BibitemShut {NoStop}%
\bibitem [{\citenamefont {Anderson}, \citenamefont {Glaser},\ and\
  \citenamefont {Glotzer}(2020)}]{anderson:cms:2020}%
  \BibitemOpen
  \bibfield  {author} {\bibinfo {author} {\bibfnamefont {J.~A.}\ \bibnamefont
  {Anderson}}, \bibinfo {author} {\bibfnamefont {J.}~\bibnamefont {Glaser}}, \
  and\ \bibinfo {author} {\bibfnamefont {S.~C.}\ \bibnamefont {Glotzer}},\
  }\href@noop {} {\bibfield  {journal} {\bibinfo  {journal} {Comput. Mater.
  Sci.}\ }\textbf {\bibinfo {volume} {173}},\ \bibinfo {pages} {109363}
  (\bibinfo {year} {2020})}\BibitemShut {NoStop}%
\bibitem [{\citenamefont {von Smoluchowski}(1916)}]{smoluchowski:pz:1916}%
  \BibitemOpen
  \bibfield  {author} {\bibinfo {author} {\bibfnamefont {M.}~\bibnamefont {von
  Smoluchowski}},\ }\href@noop {} {\bibfield  {journal} {\bibinfo  {journal}
  {Phys. Zeit.}\ }\textbf {\bibinfo {volume} {17}},\ \bibinfo {pages} {557}
  (\bibinfo {year} {1916})}\BibitemShut {NoStop}%
\bibitem [{\citenamefont {von Smoluchowski}(1918)}]{smoluchowski:zpc:1918}%
  \BibitemOpen
  \bibfield  {author} {\bibinfo {author} {\bibfnamefont {M.}~\bibnamefont {von
  Smoluchowski}},\ }\href@noop {} {\bibfield  {journal} {\bibinfo  {journal}
  {Z. Physik. Chem.}\ }\textbf {\bibinfo {volume} {92}},\ \bibinfo {pages}
  {129} (\bibinfo {year} {1918})}\BibitemShut {NoStop}%
\bibitem [{\citenamefont {Icardi}\ \emph {et~al.}(2023)\citenamefont {Icardi},
  \citenamefont {Di~Pasquale}, \citenamefont {Crevacore}, \citenamefont
  {Marchisio},\ and\ \citenamefont {Babler}}]{icardi:tpm:2023}%
  \BibitemOpen
  \bibfield  {author} {\bibinfo {author} {\bibfnamefont {M.}~\bibnamefont
  {Icardi}}, \bibinfo {author} {\bibfnamefont {N.}~\bibnamefont {Di~Pasquale}},
  \bibinfo {author} {\bibfnamefont {E.}~\bibnamefont {Crevacore}}, \bibinfo
  {author} {\bibfnamefont {D.}~\bibnamefont {Marchisio}}, \ and\ \bibinfo
  {author} {\bibfnamefont {M.~U.}\ \bibnamefont {Babler}},\ }\href@noop {}
  {\bibfield  {journal} {\bibinfo  {journal} {Transp. Porous Media}\ }\textbf
  {\bibinfo {volume} {146}},\ \bibinfo {pages} {197} (\bibinfo {year}
  {2023})}\BibitemShut {NoStop}%
\bibitem [{\citenamefont {Oles}(1992)}]{oles:jcis:1992}%
  \BibitemOpen
  \bibfield  {author} {\bibinfo {author} {\bibfnamefont {V.}~\bibnamefont
  {Oles}},\ }\href@noop {} {\bibfield  {journal} {\bibinfo  {journal} {J.
  Colloid Interface Sci.}\ }\textbf {\bibinfo {volume} {154}},\ \bibinfo
  {pages} {351} (\bibinfo {year} {1992})}\BibitemShut {NoStop}%
\bibitem [{\citenamefont {Zaccone}\ \emph {et~al.}(2011)\citenamefont
  {Zaccone}, \citenamefont {Gentili}, \citenamefont {Wu}, \citenamefont
  {Morbidelli},\ and\ \citenamefont {Del~Gado}}]{zaccone:prl:2011}%
  \BibitemOpen
  \bibfield  {author} {\bibinfo {author} {\bibfnamefont {A.}~\bibnamefont
  {Zaccone}}, \bibinfo {author} {\bibfnamefont {D.}~\bibnamefont {Gentili}},
  \bibinfo {author} {\bibfnamefont {H.}~\bibnamefont {Wu}}, \bibinfo {author}
  {\bibfnamefont {M.}~\bibnamefont {Morbidelli}}, \ and\ \bibinfo {author}
  {\bibfnamefont {E.}~\bibnamefont {Del~Gado}},\ }\href@noop {} {\bibfield
  {journal} {\bibinfo  {journal} {Phys. Rev. Lett.}\ }\textbf {\bibinfo
  {volume} {106}},\ \bibinfo {pages} {138301} (\bibinfo {year}
  {2011})}\BibitemShut {NoStop}%
\end{thebibliography}%

\end{document}